\documentclass[
    reprint,            
    amsmath,amssymb,
    aps,
    pra,
    superscriptaddress,
    nofootinbib
]{revtex4-2}

\usepackage{graphicx}
\usepackage{dcolumn}
\usepackage{bm}
\usepackage{xcolor}

\usepackage[T1]{fontenc}        
\usepackage{mathptmx}           
\usepackage[scaled=0.92]{helvet}

\setcitestyle{super}            

\usepackage[
    colorlinks=true, 
    linkcolor=blue, 
    citecolor=blue, 
    urlcolor=blue,
    breaklinks=true    
]{hyperref}

\usepackage{titlesec}

\titleformat{\section}
  {\large\bfseries\sffamily\raggedright}
  {\thesection}
  {1em}
  {}
\titlespacing*{\section}{0pt}{1.5em plus 0.5em minus 0.2em}{0.5em plus 0.2em}

\titleformat{\subsection}
  {\normalsize\bfseries\sffamily\raggedright}
  {\thesubsection}
  {1em}
  {}
\titlespacing*{\subsection}{0pt}{1.2em plus 0.5em minus 0.2em}{0.3em plus 0.2em}

\renewcommand{\figurename}{Fig.}

\makeatletter
\renewcommand*{\fnum@figure}{{\bfseries \figurename~\thefigure}}
\makeatother

\usepackage[]{hyperref}
\usepackage{color} 
\usepackage{amsmath}

\begin{document}

\title{Time-resolved correlation engineering in DLCZ Raman photon sources}

\author{Jiun-Shiuan Shiu}
\affiliation{Department of Physics, National Cheng Kung University, Tainan 70101, Taiwan}
\affiliation{Center for Quantum Frontiers of Research $\&$ Technology, Tainan 70101, Taiwan}

\author{Chang-Wei Lin}
\affiliation{Department of Physics, National Cheng Kung University, Tainan 70101, Taiwan}
\affiliation{Center for Quantum Frontiers of Research $\&$ Technology, Tainan 70101, Taiwan}

\author{Chi-Ming Yang}
\affiliation{Department of Physics, National Cheng Kung University, Tainan 70101, Taiwan}
\affiliation{Center for Quantum Frontiers of Research $\&$ Technology, Tainan 70101, Taiwan}

\author{Ite A. Yu}
\affiliation{Department of Physics and Center for Quantum Science and Technology, National Tsing Hua University, Hsinchu 30013, Taiwan}
\affiliation{National Center for Excellence in Quantum Information Science and Engineering, National Tsing Hua University, Hsinchu 30013, Taiwan}

\author{Yong-Fan Chen}
\email{yfchen@mail.ncku.edu.tw} 
\affiliation{Department of Physics, National Cheng Kung University, Tainan 70101, Taiwan}
\affiliation{Center for Quantum Frontiers of Research $\&$ Technology, Tainan 70101, Taiwan}

\date{August 13, 2026}


\begin{abstract}

Memory-assisted quantum networks require photon sources with controllable temporal and correlation properties. The Duan--Lukin--Cirac--Zoller (DLCZ) protocol provides a platform based on spontaneous Raman scattering in atomic ensembles, but a unified predictive theory connecting control parameters to correlations under realistic propagation and noise conditions remains lacking. Here we present a propagation-inclusive open-system quantum theory that retains write-induced population redistribution while combining Heisenberg--Langevin dynamics with Maxwell--Schr\"odinger propagation. We experimentally validate its key predictions. The theory predicts time-dependent Stokes generation, spin-wave evolution, retrieved anti-Stokes wavepackets, and time-resolved cross-correlations. Experiments confirm robust correlations under retrieval tuning and enhanced correlations for shorter write pulses, consistent with the different scaling of correlated coincidences and accidental backgrounds with the mean spin-wave excitation number. Classically controlled retrieval enables temporal gating and slicing of the anti-Stokes wavepacket, establishing a quantitative framework for correlation engineering in memory-compatible DLCZ photon sources.

\end{abstract}


\maketitle


\newcommand{\FigOne}{
    \begin{figure}[t]
    \centering
    \includegraphics[width = 8.6 cm]{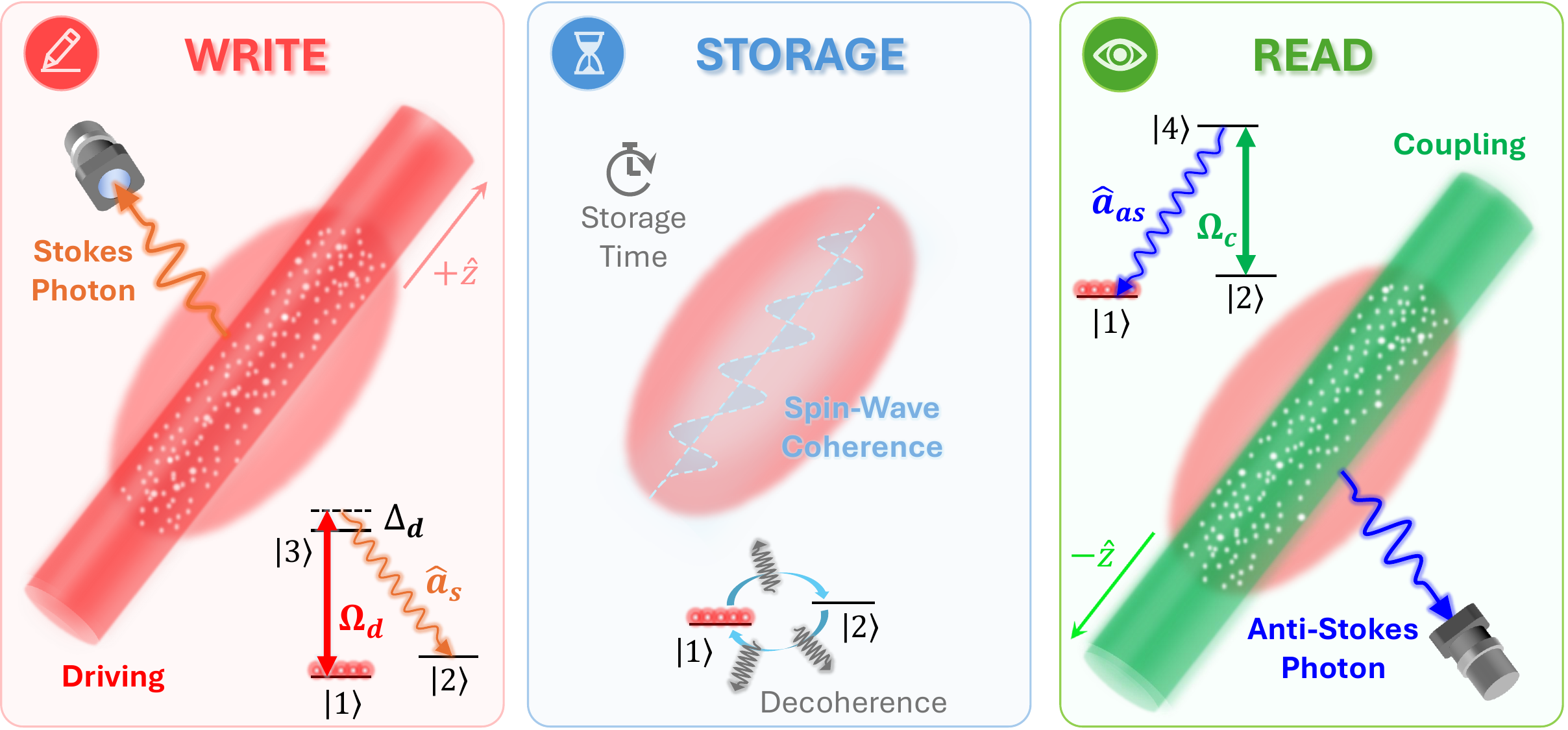}
    \caption{
\textbf{Schematic of the DLCZ-type SRS process.}
During the write stage, a driving pulse with Rabi frequency $\Omega_d$ and detuning $\Delta_d$ excites the $|1\rangle\rightarrow|3\rangle$ transition. Stokes emission on $|3\rangle\rightarrow|2\rangle$ creates a collective $|1\rangle$--$|2\rangle$ spin wave, which is stored for a controllable interval and decays through ground-state decoherence. During readout, a coupling pulse with Rabi frequency $\Omega_c$ maps the stored spin wave onto an anti-Stokes field via state $|4\rangle$. The Stokes and anti-Stokes fields propagate along $+z$ and $-z$, respectively.
}
    \label{fig1}
    \end{figure}
}

\newcommand{\FigTwo}{
    \begin{figure}[t]
    \centering
    \includegraphics[width = 8.8 cm]{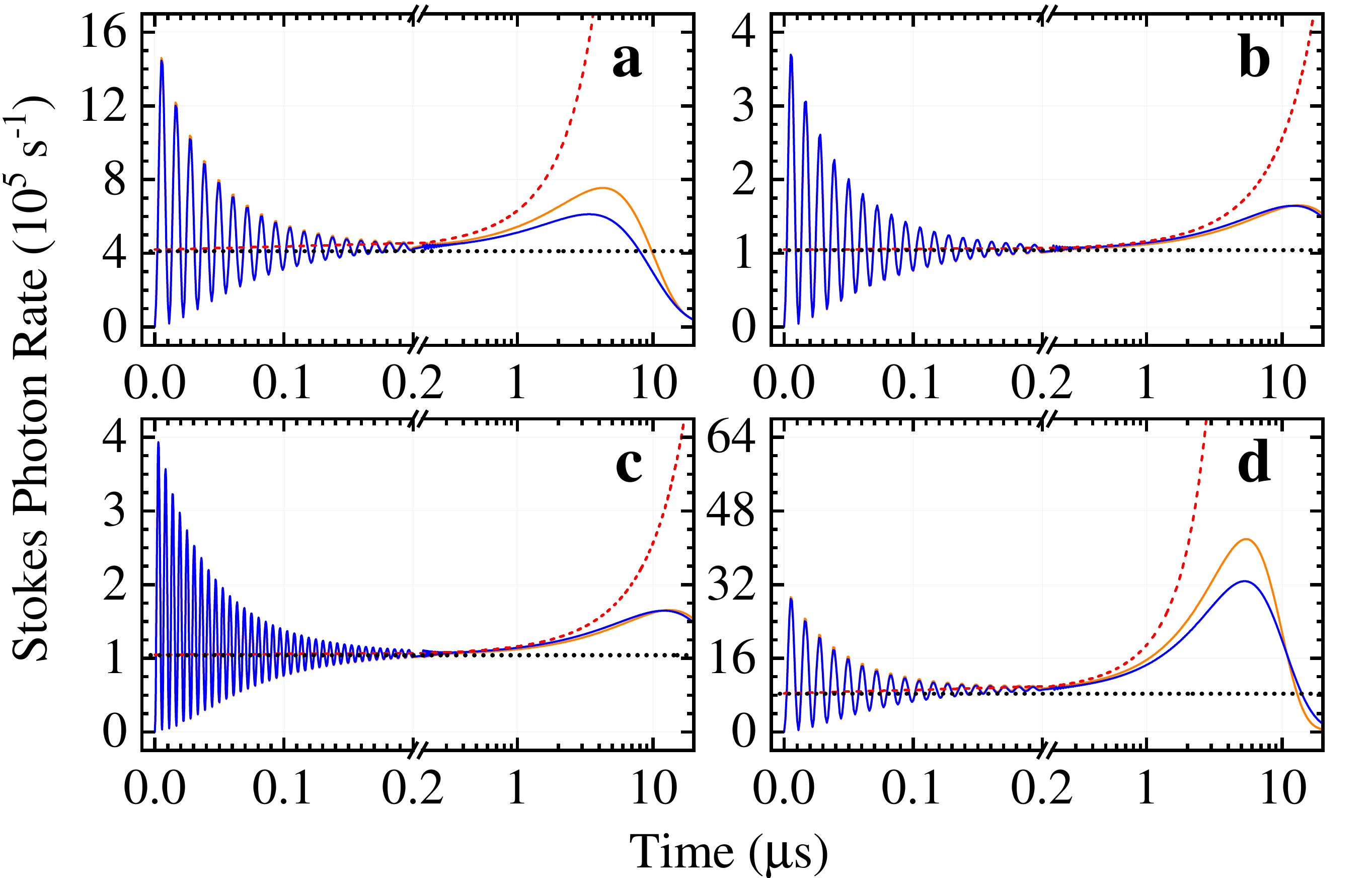}
    \caption{
\textbf{Comparison of Stokes photon generation rates in DLCZ-type SRS and SFWM.}
Panels \textbf{a}--\textbf{d} use $(\mathrm{OD},\Omega_d,\Delta_d)=(10,2\Gamma,15\Gamma)$, $(10,1\Gamma,15\Gamma)$, $(10,2\Gamma,30\Gamma)$, and $(20,2\Gamma,15\Gamma)$, respectively, with $\gamma_{21}=0.001\Gamma$. The orange solid curves show the time-dependent open-system calculations using the piecewise-constant propagator. The blue solid curves show phenomenological depletion-corrected large-detuning estimates obtained by substituting $\mathrm{OD}_{\mathrm{eff}}(T)=\mathrm{OD}\langle\hat{\sigma}_{11}(T)\rangle$ into the undepleted response of Eq.~\eqref{eq15}. The red dashed curves show the undepleted-medium results of Eq.~\eqref{eq16}, and the black dotted curves show the steady-state SFWM predictions under identical driving conditions~\cite{Kolchin2007}.
}
    \label{fig2}
    \end{figure}
}

\newcommand{\FigThree}{
    \begin{figure}[t]
    \centering
    \includegraphics[width = 8.8 cm]{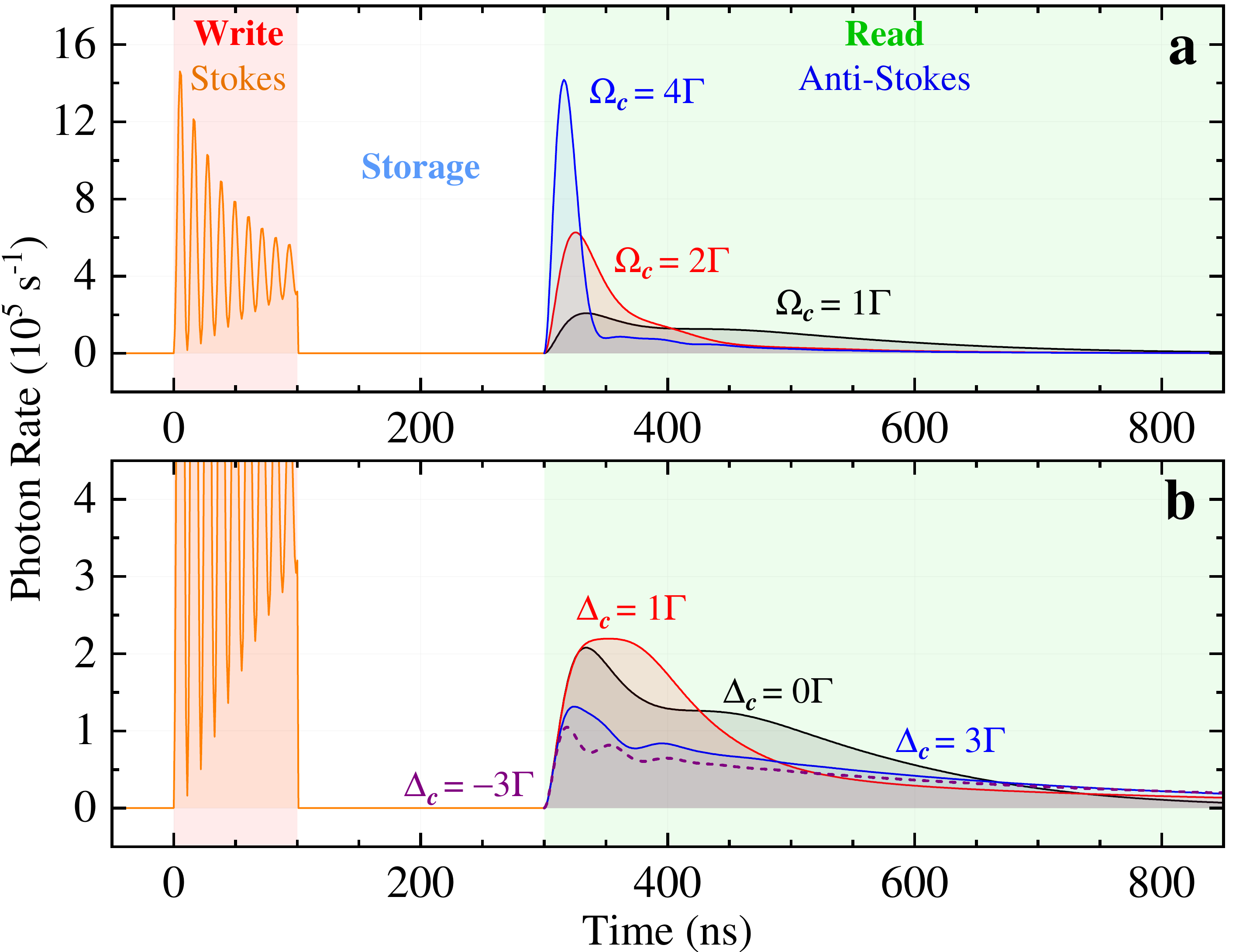}
    \caption{
\textbf{Temporal profiles of Stokes generation and anti-Stokes retrieval in DLCZ-type SRS.}
Panel \textbf{a} shows the retrieval dynamics for different coupling strengths $\Omega_c$ at $\Delta_c=0$. Panel \textbf{b} shows the retrieval dynamics for different coupling detunings $\Delta_c$ at $\Omega_c=1\Gamma$. In both panels, the driving parameters are $\Omega_d=2\Gamma$ and $\Delta_d=15\Gamma$, with $\mathrm{OD}=10$, $\gamma_{21}=0.001\Gamma$, and $\Delta kL=0.37\pi$. A 100-ns driving pulse is followed by a 200-ns storage interval before retrieval. The orange solid curves represent the generated Stokes wavepackets, while the remaining curves represent the retrieved anti-Stokes wavepackets.
}
    \label{fig3}
    \end{figure}
}

\newcommand{\FigFour}{
    \begin{figure}[t]
    \centering
    \includegraphics[width = 8.8 cm]{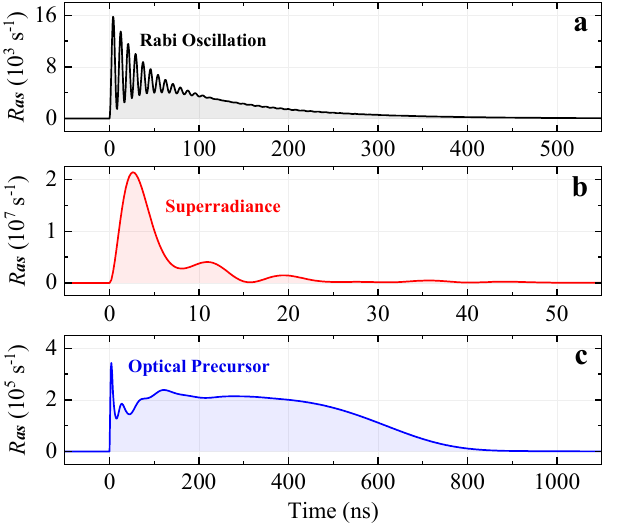}
    \caption{
\textbf{Anti-Stokes retrieval dynamics in DLCZ-type SRS under different regimes.}
Panel \textbf{a} shows Rabi-like retrieval at $\mathrm{OD}=1$, $\Omega_c=20\Gamma$, and $\Delta_c=0$. Panel \textbf{b} shows superradiant retrieval at $\mathrm{OD}=100$, $\Omega_c=20\Gamma$, and $\Delta_c=0$. Panel \textbf{c} shows an optical precursor followed by the main retrieval wavepacket at $\mathrm{OD}=100$, $\Omega_c=2\Gamma$, and $\Delta_c=0$. In all panels, the driving parameters are $\Omega_d=2\Gamma$ and $\Delta_d=30\Gamma$, and the calculations use $\gamma_{21}=0$ and $\Delta kL=0.37\pi$.
}
    \label{fig4}
    \end{figure}
}

\newcommand{\FigFive}{
    \begin{figure}[t]
    \centering
    \includegraphics[width = 8.8 cm]{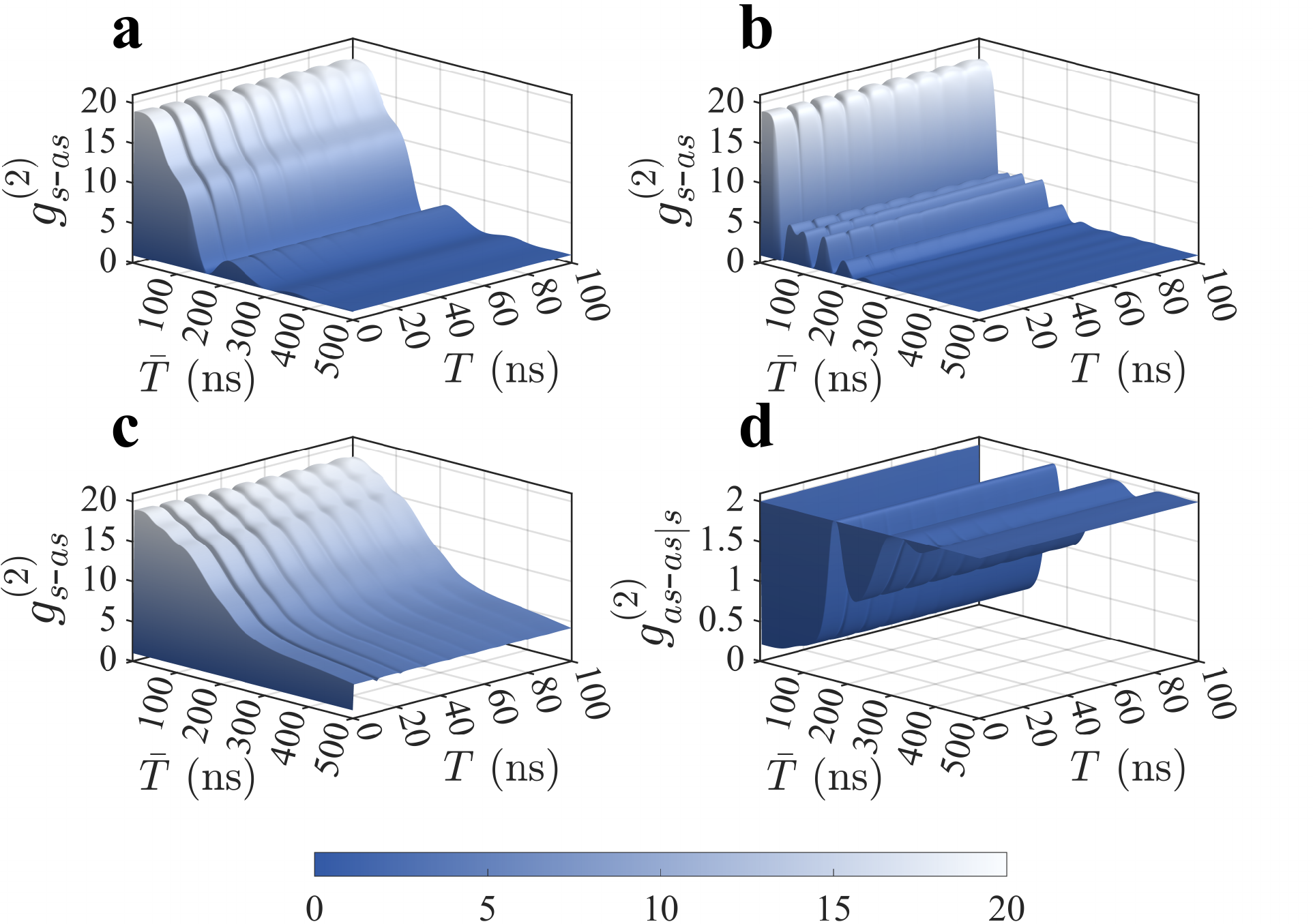}
    \caption{
\textbf{Two-photon correlation functions in DLCZ-type SRS under different coupling conditions.}
Panels \textbf{a}--\textbf{c} show the calculated time-resolved normalized cross-correlation profiles $g_{s\text{-}as}^{(2)}(T,\bar{T})$. The retrieval parameters $(\Omega_c,\Delta_c)$ are $(2\Gamma,0)$, $(4\Gamma,0)$, and $(2\Gamma,2\Gamma)$ in panels \textbf{a}, \textbf{b}, and \textbf{c}, respectively. Panel \textbf{d} shows the corresponding model-inferred time-resolved zero-delay conditional anti-Stokes autocorrelation under the retrieval conditions of panel \textbf{a}. The quantities in all panels are not averaged over finite detection windows. The calculations use a 100-ns driving pulse followed by a 200-ns storage interval. The driving parameters are fixed at $\Omega_d=2\Gamma$ and $\Delta_d=15\Gamma$, with $\mathrm{OD}=10$, $\gamma_{21}=0.001\Gamma$, and $\Delta kL=0.37\pi$.
}
    \label{fig5}
    \end{figure}
}

\newcommand{\FigSix}{
	\begin{figure}[t]
	\centering
	\includegraphics[width = 8.8 cm]{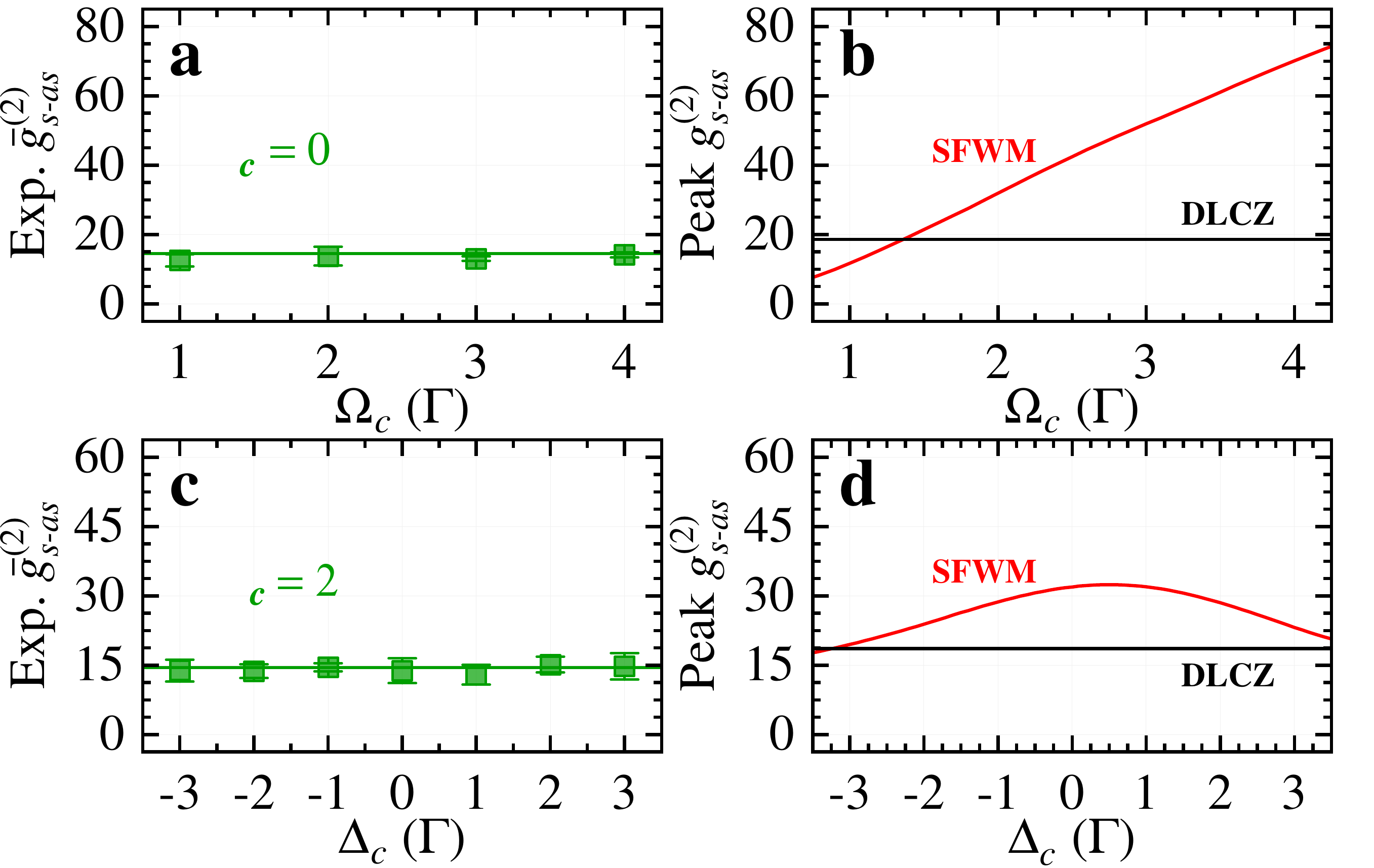}
	\caption{
\textbf{Two-photon correlations in DLCZ-type SRS and comparison with SFWM under different coupling conditions.}
Panels \textbf{a} and \textbf{c} show the measured integrated cross-correlations versus $\Omega_c$ at $\Delta_c=0\Gamma$ and versus $\Delta_c$ at $\Omega_c=2\Gamma$, respectively. Squares represent the experimental data, and the solid curves represent the theoretical integrated cross-correlations. Panels \textbf{b} and \textbf{d} show the corresponding theoretical time-resolved peak cross-correlations for DLCZ-type SRS and SFWM. The integrated measurements and theoretical peak values are distinct temporal observables and are not expected to coincide quantitatively. Unless otherwise stated, all experimental error bars throughout this work represent one standard deviation of the statistical uncertainty. The parameters are $\mathrm{OD}=10$, $\Omega_d=2\Gamma$, $\Delta_d=15\Gamma$, $\gamma_{21}=0.001\Gamma$, and $\Delta kL=0.37\pi$.
}
	\label{fig6}
	\end{figure}
}

\newcommand{\FigSeven}{
    \begin{figure}[t]
    \centering
    \includegraphics[width = 8.8 cm]{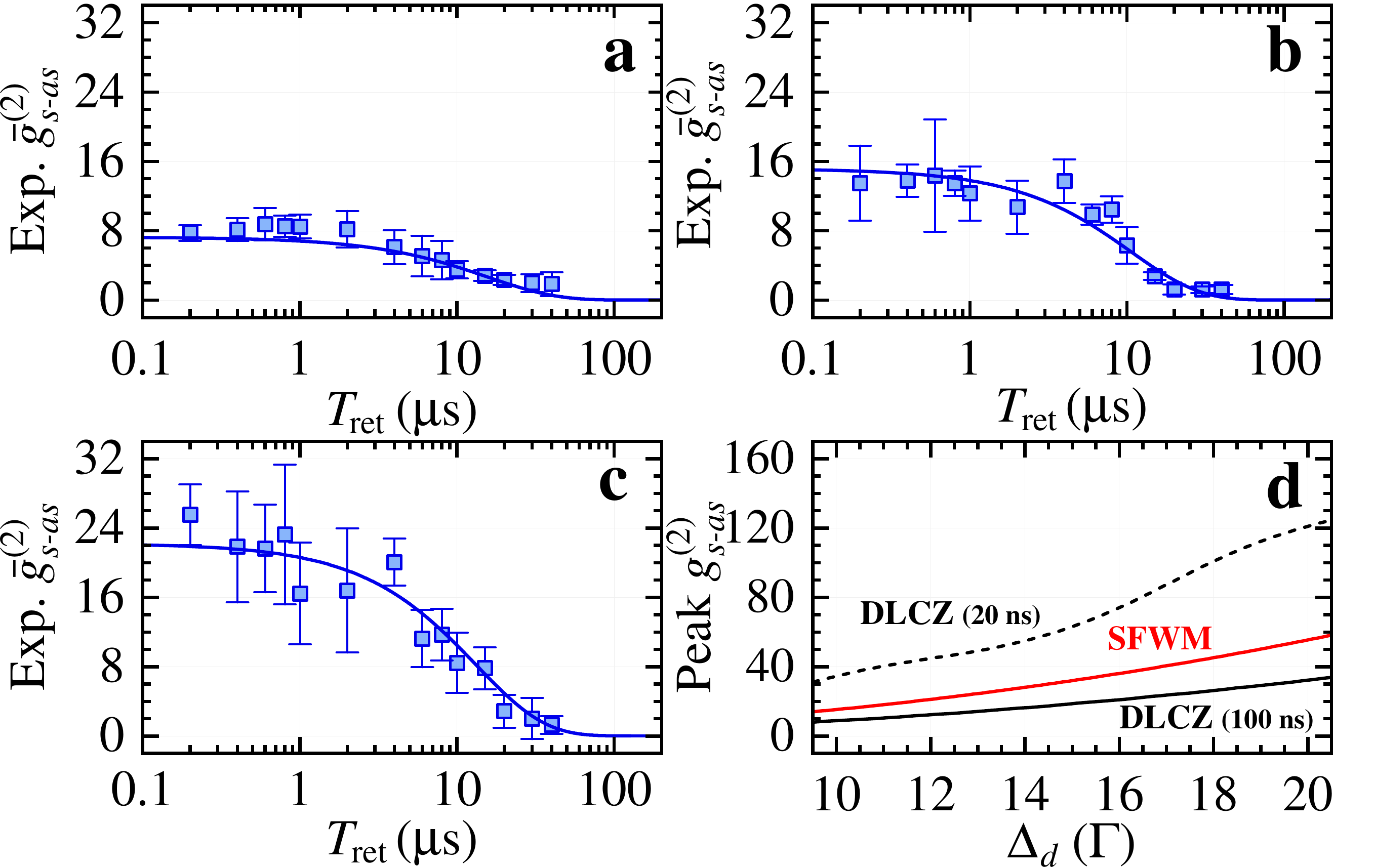}
    \caption{
\textbf{Two-photon correlations in DLCZ-type SRS and comparison with SFWM as a function of driving detuning.} Panels \textbf{a}--\textbf{c} show the measured integrated cross-correlations as functions of retrieval time $T_{\rm ret}$ for driving detunings $\Delta_d=10\Gamma$, $15\Gamma$, and $20\Gamma$, respectively. Squares represent experimental data, and solid curves represent exponential fits. Panel \textbf{d} shows the theoretical peak cross-correlation values as functions of $\Delta_d$ for DLCZ-type SRS with two driving pulse durations, $T_d=100$ ns and $20$ ns, together with the corresponding SFWM results evaluated under otherwise comparable operating conditions. In all panels, the parameters are ${\rm OD}=10$, $\Omega_d=\Omega_c=2\Gamma$, $\gamma_{21}=0.001\Gamma$, and $\Delta kL=0.37\pi$. 
}
    \label{fig7}
    \end{figure}
}

\newcommand{\FigEight}{
	\begin{figure}[t]
	\centering
	\includegraphics[width = 8.6 cm]{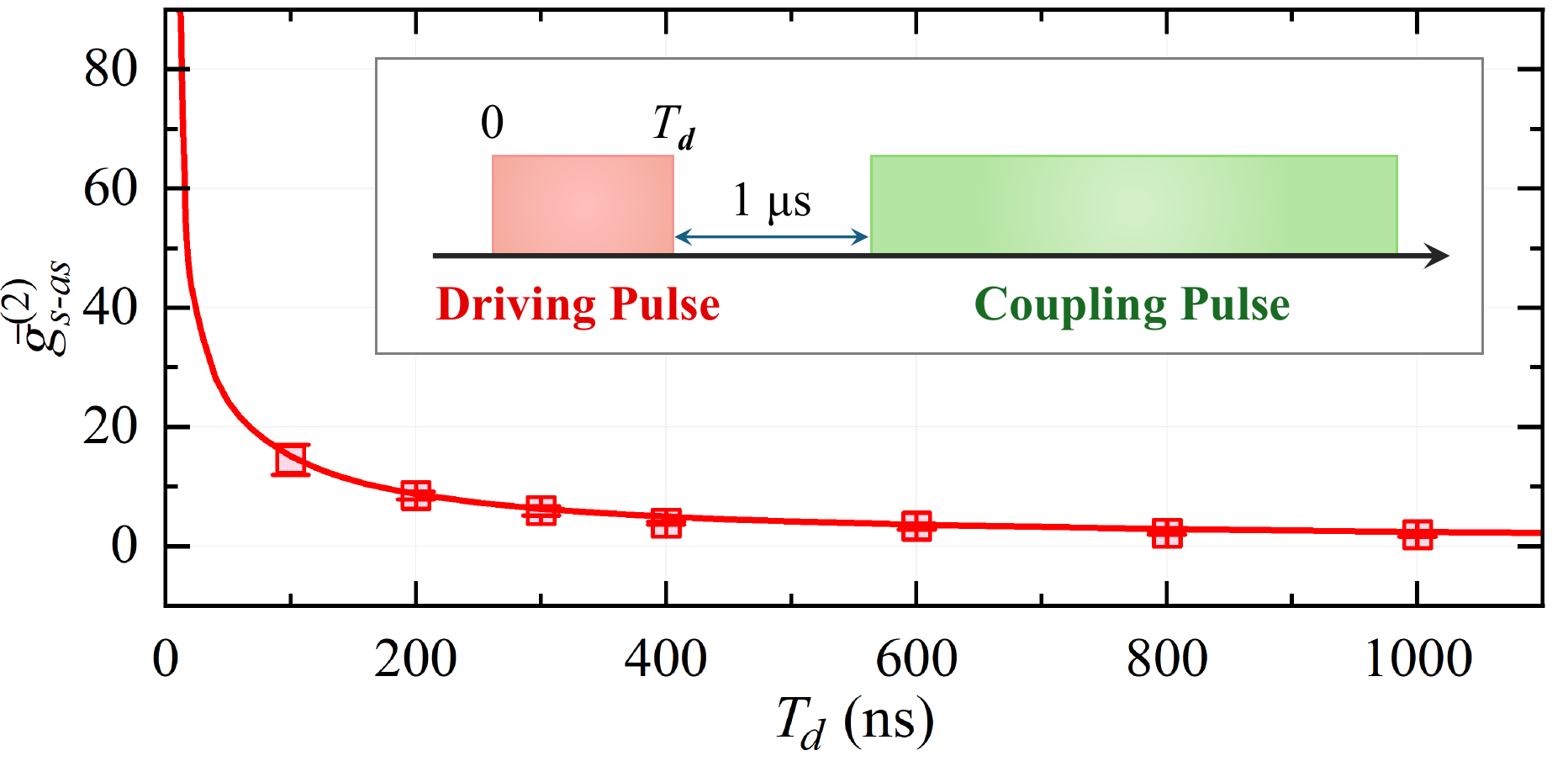}
	\caption{
\textbf{Two-photon correlations in DLCZ-type SRS as a function of driving pulse duration.}
Squares represent the experimental data, and the solid curve represents the theoretical integrated cross-correlation evaluated using the same temporal integration procedure. The correlation decreases with increasing driving pulse duration, showing that shorter write pulses enhance the correlation by reducing the relative accidental-coincidence contribution. The measurements are performed with a storage time of 1~$\mu$s. The parameters are $\mathrm{OD}=10$, $\Omega_d=\Omega_c=2\Gamma$, $\Delta_d=15\Gamma$, $\gamma_{21}=0.001\Gamma$, and $\Delta kL=0.37\pi$.
}
	\label{fig8}
	\end{figure}
}

\newcommand{\FigNine}{
	\begin{figure}[t]
	\centering
	\includegraphics[width = 8.5 cm]{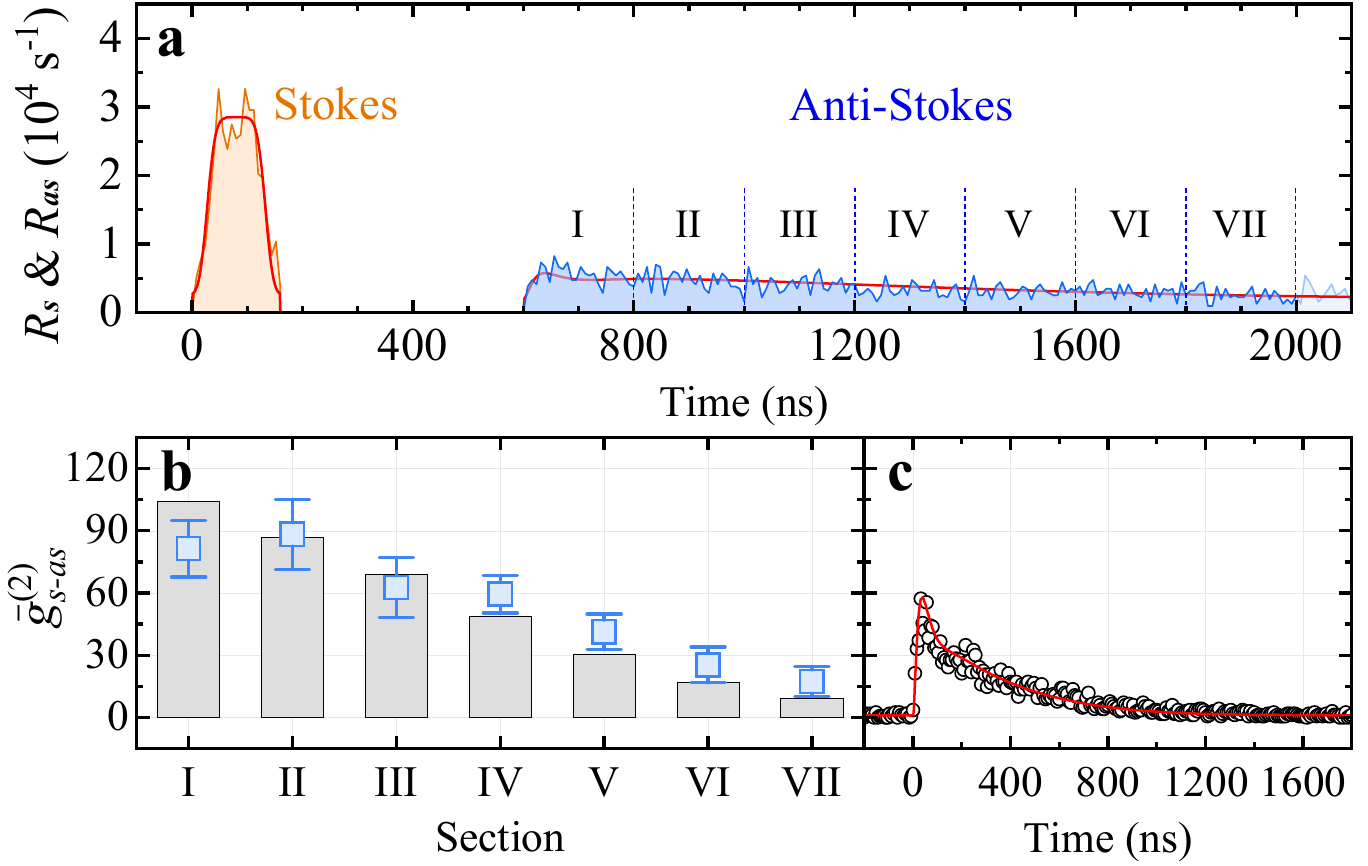}
	\caption{
\textbf{Temporal slicing analysis of biphoton correlations in DLCZ-type SRS.}
Panel \textbf{a} shows the measured Stokes and anti-Stokes temporal profiles; the orange and red solid curves show the calculated profiles, respectively. The calculation includes the finite turn-on of the acousto-optically switched write field. The retrieved anti-Stokes wavepacket is divided into seven 200-ns sections (I--VII). The parameters are $\mathrm{OD}=10$, $\Omega_d=\Omega_c=0.5\Gamma$, $\Delta_d=15\Gamma$, $\gamma_{21}=0.001\Gamma$, and $\Delta kL=0.37\pi$. Panel \textbf{b} shows the measured cross-correlations within the anti-Stokes sections (blue squares), with the theoretical predictions (gray bars) evaluated using the same integration windows and including leakage-light and detector-dark-count backgrounds. Panel \textbf{c} shows the time-resolved SFWM cross-correlations using an 8-ns integration window under otherwise identical experimental parameters, with temporally overlapping driving and coupling pulses. Circles and the red solid curve represent experiment and theory, respectively.
}
	\label{fig9}
	\end{figure}
}

\newcommand{\FigTen}{
    \begin{figure*}[t]
    \centering
    \includegraphics[width = 17.8 cm]{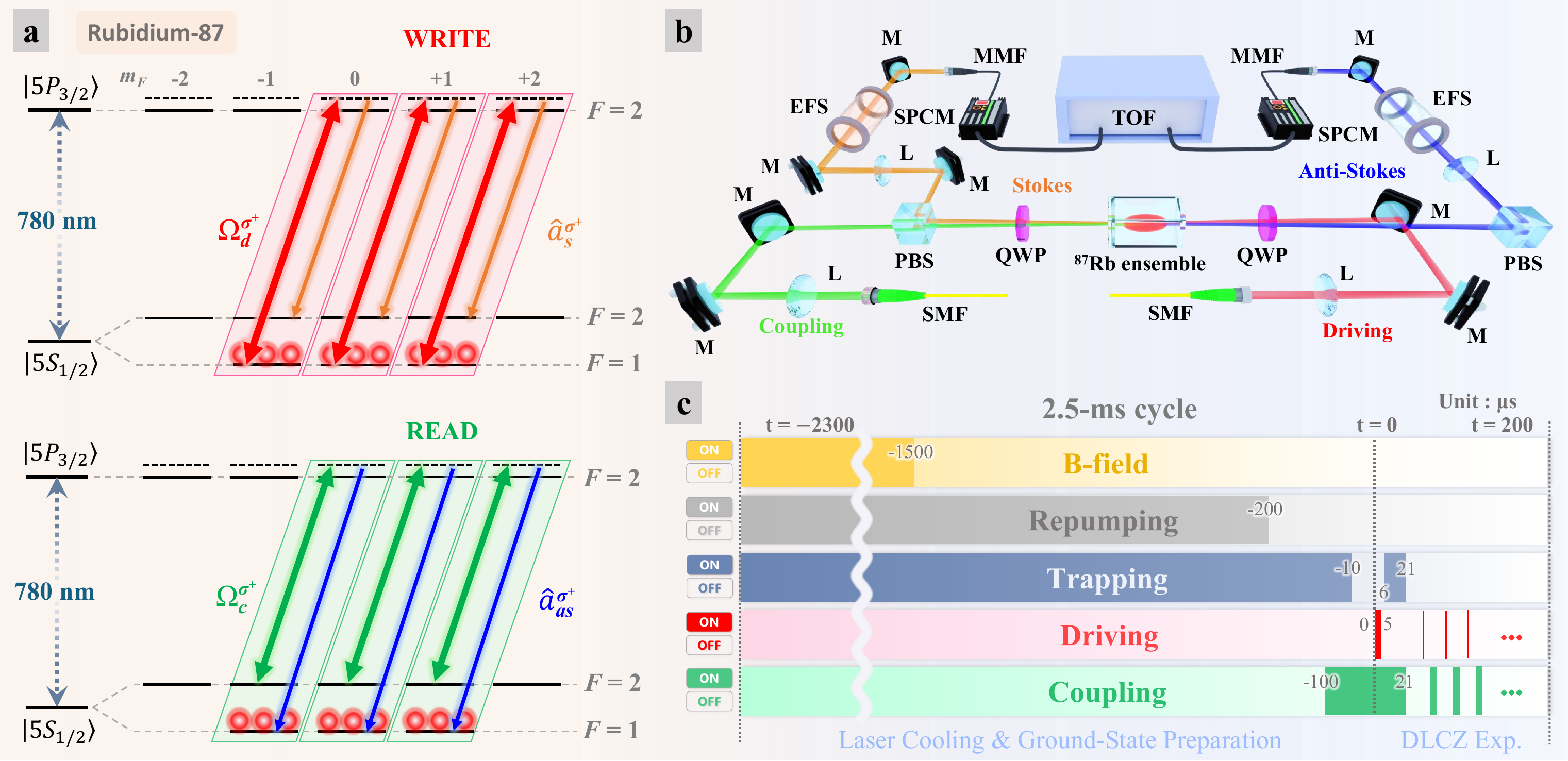}
    \caption{
\textbf{Experimental implementation of DLCZ-type SRS.}
\textbf{a} Energy-level diagrams for the write and read processes. \textbf{b} Schematic of the optical setup. EFS, etalon filter set; L, lens; M, mirror; MMF, multimode fiber; PBS, polarizing beam splitter; QWP, quarter-wave plate; SMF, single-mode fiber; SPCM, single-photon counting module; TOF, time-of-flight multiscaler. \textbf{c} Experimental timing sequence within each 2.5-ms cycle, including laser cooling, in situ SFWM-based OD calibration, optical pumping, and DLCZ-type SRS measurement.
}
    \label{fig10}
    \end{figure*}
}

\section*{INTRODUCTION} \label{sec:Introduction} 

Photon pairs are a central resource for quantum communication and distributed quantum networks~\cite{Kimble2008,Wehner2018,Marcikic2003,Yin2017}. For memory-assisted architectures, narrowband, time-correlated photon pairs are essential for coupling to quantum memories and synchronizing remote nodes~\cite{commun1,commun2,commun3,commun4,memory1,memory2,memory3,memory4,memory5,memory6}. Atomic ensembles are attractive because their optical transitions naturally produce narrowband emission with well-defined temporal structure, enabling efficient light--matter interfaces for networked quantum systems~\cite{Hammerer2010}. A canonical paradigm is the Duan--Lukin--Cirac--Zoller (DLCZ) protocol~\cite{DLCZ}. It relies on probabilistic Stokes-photon detection to herald a collective spin-wave excitation, followed by on-demand retrieval into an anti-Stokes photon. Here we call the underlying write process DLCZ-type spontaneous Raman scattering (SRS). A write pulse drives a Raman transition that emits a Stokes photon while creating a collective ground-state coherence. A read pulse then converts this spin-wave coherence into a directed anti-Stokes field. This write--store--read sequence provides a memory-assisted photon source compatible with atomic quantum memories and quantum repeater architectures.

Over the past two decades, DLCZ-type SRS has been extensively investigated experimentally. Key advances include nonclassical photon-pair generation and heralded single-photon operation~\cite{g1,g2,g3,g4,g5,g6,g7}, atom--photon and photon--photon entanglement~\cite{pol1,pol2,pol3,pol4,pol5,pol6,pol7}, long-lived spin-wave storage and controlled photon retrieval~\cite{long1,long2,long3,long4,long5,long6,long7,long8,long9,long10}, and high retrieval efficiency~\cite{ret1,ret2,ret3,ret4,ret5,ret6} in both cold and thermal ensembles. Beyond single-node operation, multi-ensemble architectures have enabled quantum-state transfer, distributed entanglement, and elementary quantum-network links~\cite{2ensemble1,2ensemble2,2ensemble3,2ensemble4,2ensemble5,2ensemble7,2ensemble8,2ensemble9,2ensemble10}. Interference between photons retrieved from independent memories, including Hong--Ou--Mandel interference, provides a sensitive benchmark of photon indistinguishability and temporal-mode matching~\cite{HOM1,HOM2,HOM3}. Multiplexed memory architectures have increased the number of accessible quantum modes across individually addressable memory cells~\cite{multiens1,multiens2,multiens3,multiens4,multiens5,multiens6}, spatial modes~\cite{Smulti1,Smulti2,Smulti3,Smulti4}, and temporal modes~\cite{Tmulti1,Tmulti2,Tmulti4,Tmulti5}. Together with entanglement swapping~\cite{ES1,ES2,ES3}, these developments have advanced scalable memory-based quantum networks.

Foundational theories established quantum-initiated Raman propagation in longitudinal and transverse geometries under undepleted-medium approximations~\cite{Raymer,Mostowski}. Later three-dimensional work extended this treatment to finite geometries and multimode effects~\cite{Sorensen2009}. Previous theories have described photon-number evolution, photon statistics, and Stokes--anti-Stokes correlations in DLCZ-type systems under weak-excitation, spin-wave, or effective photon-pair approximations~\cite{DLCZ,Sisakyan2005,Sangouard2011}. Short-pulse spontaneous Raman generation has also been optimized in cold atomic ensembles~\cite{Ho2018}, while different initial populations were considered in a one-dimensional SRS model~\cite{Ooi2024}. Related $\Lambda$-type theories and experiments have clarified spin-wave storage, retrieval efficiency, and photon-wavepacket shaping~\cite{Gorshkov2007PRA1,Gorshkov2007PRA2,long6}. Despite these advances, a unified propagation-inclusive open-system framework that retains the time-dependent write-driven populations and directly predicts photon rates, retrieved wavepackets, and time-resolved correlations under experimental conditions remains lacking. This limitation is important for extended write pulses, where optical pumping transfers population from the initial ground state, suppresses transient Raman gain, and modifies subsequent correlations. Connecting experimental controls to measured correlations therefore often relies on limiting models or empirical optimization.

A related challenge is that DLCZ-type SRS and double-$\Lambda$ spontaneous four-wave mixing (SFWM)~\cite{SFWM1,SFWM2,SFWM3,SFWM4,SFWM5} rely on closely related atomic level configurations, yet they are commonly analyzed using different theoretical approximations. For SFWM, open-system quantum Langevin descriptions are well established~\cite{Kolchin2007,SFWM6} and have been extended to increasingly complex dynamical regimes~\cite{SFWM7,SFWM8,SFWM9,SFWM10}. An equally explicit propagation-inclusive treatment of the sequential DLCZ write, storage, and read processes remains less developed, particularly when spontaneous-emission noise and time-dependent population redistribution are retained within the same formalism. This has hindered systematic comparisons of photon-generation rates, retrieval wavepackets, accidental backgrounds, and two-photon correlations in DLCZ-type SRS and SFWM.

\FigOne

Here we establish a predictive open-system description of DLCZ-type SRS that provides a direct quantitative connection between control parameters, spatiotemporal wavepackets, and two-photon correlations. The theory retains the full time dependence of the write-driven atomic populations and resolves the crossover from transient Raman enhancement to population-transfer-induced suppression. By combining these atomic dynamics with coupled Heisenberg--Langevin and Maxwell--Schr\"odinger equations, it predicts Stokes generation beyond conventional undepleted-population and adiabatic approximations. In the strict undepleted large-detuning limit, the theory also yields compact analytic response functions, while a phenomenological replacement of the optical depth (OD) by the time-dependent quantity $\mathrm{OD}_{\mathrm{eff}}(T)$ captures the leading depletion effect. The same framework predicts the retrieved anti-Stokes wavepacket and the full time-resolved normalized cross-correlation function, enabling direct quantitative comparison with experimental measurements.

Within this open-system framework, we identify a physically transparent and experimentally testable distinction between DLCZ-type SRS and double-$\Lambda$ SFWM. In the DLCZ scheme, the normalized Stokes--anti-Stokes correlation remains nearly unchanged over the range of retrieval bandwidth and frequency tuning considered here. This behavior arises because retrieval parameters modify both the conditional readout signal and the accidental background in a correlated manner. Measurements confirm the weak dependence of the correlation on retrieval tuning and are consistent with the predicted correlated variation of these contributions. We further show that shorter write pulses enhance the two-photon correlation because the correlated coincidences and accidental backgrounds exhibit different scaling with the mean number of spin-wave excitations. Finally, we experimentally demonstrate temporal slicing of the retrieved anti-Stokes wavepacket and evaluate the corresponding Stokes--anti-Stokes correlations in individual temporal sections. This measurement shows how the classically controlled readout stage of the DLCZ protocol enables direct temporal gating and selection of the retrieved wavepacket without requiring real-time feed-forward from Stokes-photon detection.


\section*{RESULTS} \label{sec:Theo}

The DLCZ-type SRS process considered here is schematically shown in Fig.~\ref{fig1}. The interaction consists of three sequential stages. During the write stage, a driving field induces SRS, with the emission of a Stokes photon accompanied by the creation of a collective ground-state spin-wave coherence in the atomic ensemble. This coherence is stored for a controllable interval and decays due to ground-state decoherence. During the read stage, a counter-propagating coupling pulse coherently maps the stored spin-wave excitation onto a phase-matched anti-Stokes field through the $|2\rangle\rightarrow|4\rangle\rightarrow|1\rangle$ read pathway, completing the write--store--read sequence.

To connect control parameters with experimentally measurable observables, including Stokes and anti-Stokes wavepackets and two-photon correlations, we model the system using a propagation-inclusive open-system quantum description based on coupled Heisenberg--Langevin equations (HLEs) and Maxwell--Schr\"odinger equations~\cite{Kolchin2007,JS3}. This framework explicitly retains spontaneous emission, Langevin noise, and spatiotemporal propagation during both the write and read processes. During the write process, the mean atomic populations and optical coherence retain their full time dependence, while the generated Stokes field and the associated atomic coherences are treated to first order.

The Stokes ($l=s$) and anti-Stokes ($l=as$) fields are treated quantum mechanically as $\vec{E}_l=\hat{e}_l\mathcal{E}_l\hat{a}_l(z,t) e^{i(\vec{k}_l\cdot\vec{r}-\omega_l t)}+\mathrm{h.c.}$, where $\mathcal{E}_l=\sqrt{\hbar\omega_l/(2\epsilon_0V)}$ is the single-photon field amplitude in the quantization volume $V$. The corresponding atom--field coupling strengths are $g_s=d_{32}\mathcal{E}_s/\hbar$ and $g_{as}=d_{41}\mathcal{E}_{as}/\hbar$, where $d_{32}$ and $d_{41}$ are the dipole matrix elements projected onto the corresponding field polarizations. The collective atomic dynamics are described by slowly varying operators $\hat{\sigma}_{jk}(z,t)$, defined as local averages over thin slices of thickness $\Delta z$. Specifically, $\hat{\sigma}_{jk}(z,t)=N_z^{-1}\sum_{m=1}^{N_z}\hat{\sigma}_{jk}^{(m)}(t)$, with $N_z=N\Delta z/L$, where $N$ is the total number of atoms and $L$ is the length of the ensemble.


\subsection*{The write-stage dynamics}

In the write stage, a classical driving field induces a Raman interaction in which Stokes-photon emission creates a collective ground-state spin-wave coherence. Under the rotating-wave approximation, this process is described by the interaction Hamiltonian
\begin{equation}
	\hat{H}_{\rm W}
	=
	-\frac{\hslash N}{2L}
	\int_0^L dz
	\bigl(
	\Delta_d \hat{\sigma}_{33}
	+
	\Omega_d \hat{\sigma}_{31}
	+
	2 g_s \hat{a}_s \hat{\sigma}_{32}
	+
	\mathrm{h.c.}
	\bigr),
\end{equation}
where $\Omega_d$ is the complex Rabi frequency and $\Delta_d$ is the detuning of the driving field. From this Hamiltonian, we obtain the HLEs for the coupled atomic populations, optical and spin coherences, and Langevin fluctuations,
\begin{align}
\frac{\partial\hat{\sigma}_{11}}{\partial t}
=&
\frac{i}{2}(\Omega_d^*\hat{\sigma}_{13}-\Omega_d\hat{\sigma}_{31})
+\Gamma_{31}\hat{\sigma}_{33}+\hat{F}_{11}
,\label{eq2}\\
\frac{\partial\hat{\sigma}_{22}}{\partial t}
=&
i(g_s^*\hat{a}_s^\dagger\hat{\sigma}_{23}-g_s\hat{a}_s\hat{\sigma}_{32})
+\Gamma_{32}\hat{\sigma}_{33}+\hat{F}_{22}
,\label{eq3}\\
\frac{\partial\hat{\sigma}_{33}}{\partial t}
=&
\frac{i}{2}(\Omega_d\hat{\sigma}_{31}-\Omega_d^*\hat{\sigma}_{13})
+i(g_s\hat{a}_s\hat{\sigma}_{32}-g_s^*\hat{a}_s^\dagger\hat{\sigma}_{23})
\nonumber\\&
-\Gamma_3\hat{\sigma}_{33}+\hat{F}_{33}
,\label{eq4}\\
\frac{\partial\hat{\sigma}_{31}}{\partial t}
=&
\frac{i}{2}\Omega_d^*(\hat{\sigma}_{33}-\hat{\sigma}_{11})
-ig_s^*\hat{a}_s^\dagger\hat{\sigma}_{21}
-\frac{\gamma_{13}^*}{2}\hat{\sigma}_{31}+\hat{F}_{31}
,\label{eq5}\\
\frac{\partial\hat{\sigma}_{23}}{\partial t}
=&
\frac{i}{2}\Omega_d\hat{\sigma}_{21}
+ig_s\hat{a}_s(\hat{\sigma}_{22}-\hat{\sigma}_{33})
-\frac{\gamma_{23}}{2}\hat{\sigma}_{23}+\hat{F}_{23}
,\label{eq6}\\
\frac{\partial\hat{\sigma}_{21}}{\partial t}
=&
\frac{i}{2}\Omega_d^*\hat{\sigma}_{23}
-ig_s\hat{a}_s\hat{\sigma}_{31}
-\frac{\gamma_{21}}{2}\hat{\sigma}_{21}+\hat{F}_{21}
.\label{eq7}
\end{align}
The complex optical-coherence decay parameters are given by $\gamma_{13}=\gamma_{23}=\Gamma_3-2i\Delta_d$, where $\Gamma_3=\Gamma_{31}+\Gamma_{32}$ is the total spontaneous decay rate of the excited state $|3\rangle$. The parameter $\gamma_{21}$ denotes the ground-state spin-wave decoherence rate. The operators $\hat{F}_{jk}$ are zero-mean Langevin noise operators that account for quantum fluctuations associated with spontaneous emission and other dissipative processes. Under the Markovian approximation, they are delta-correlated in space and time. We retain the full time dependence of the mean atomic populations and optical coherence driven by the classical write field. The generated Stokes field and the associated Raman coherences are treated to first order, so their backaction on the mean atomic dynamics is neglected. This perturbative ordering does not impose an undepleted-population approximation, since write-induced population redistribution remains fully included.

We first solve the resulting time-dependent mean-value equations to obtain $\langle\hat{\sigma}_{11}(t)\rangle$, $\langle\hat{\sigma}_{22}(t)\rangle$, $\langle\hat{\sigma}_{33}(t)\rangle$, and $\langle\hat{\sigma}_{31}(t)\rangle$. These quantities determine the coefficients governing the Stokes polarization $\hat{\sigma}_{23}$ and spin-wave coherence $\hat{\sigma}_{21}$ in Eqs.~\eqref{eq6} and \eqref{eq7}. The additive Langevin sources in these equations are retained to describe the quantum initiation of the Raman process. The propagation of the Stokes field is described by the Maxwell--Schr\"odinger equation (MSE)
\begin{align}
\left(\frac{\partial}{\partial z}+\frac{1}{c}\frac{\partial}{\partial t}\right)\hat{a}_s=\frac{ig_s^*N}{c}\hat{\sigma}_{23}
,\label{eq8}
\end{align}
which closes the coupled light--matter dynamics by linking the field evolution to the atomic polarization.

To solve the coupled equations, we transform to a moving frame defined by $Z=z$ and $T=t-z/c$, for which the traveling-wave operator $\partial_z+c^{-1}\partial_t$ reduces to $\partial_Z$ at fixed $T$. We then apply the spatial Laplace transform $\widetilde{f}(s,T)=\int_0^\infty dZ\,e^{-sZ}f(Z,T)$, which converts the propagation equation into an algebraic relation. Although the transform is defined over the positive spatial half-axis, the atomic operators and Langevin sources vanish outside the medium $0\leq Z\leq L$. The physical inverse-transformed solutions are evaluated within the ensemble and at the output boundary $Z=L$.

In the Laplace domain, the collective atomic and Langevin noise operators are denoted by $\hat{S}_{jk}(s,T)=\mathcal{L}[\hat{\sigma}_{jk}(Z,T)]$ and $\hat{G}_{jk}(s,T)=\mathcal{L}[\hat{F}_{jk}(Z,T)]$, respectively, and the Stokes field by $\hat{A}_s(s,T)=\mathcal{L}[\hat{a}_s(Z,T)]$. Applying this transformation to Eq.~\eqref{eq8} gives $\hat{A}_s(s,T)=s^{-1}\hat{a}_s(0,T)+(ig_s^*N/cs)\hat{S}_{23}(s,T)$. Substitution of this relation into the atomic equations gives the coupled dynamics of $\hat{S}_{23}$ and $\hat{S}_{21}$ in the $s$ domain as
\begin{align}
	\frac{\partial\mathbf{S}(s,T)}{\partial T}=\mathbf{M}(s,T)\mathbf{S}(s,T)+\mathbf{N}(s,T).
	\label{eq9}
\end{align}
Here, $\mathbf{S}(s,T)=\begin{bmatrix}\hat{S}_{23}(s,T),\hat{S}_{21}(s,T)\end{bmatrix}^{\rm T}$ contains the Stokes polarization $\hat{S}_{23}$ and the spin-wave coherence $\hat{S}_{21}$. The elements of the coupling matrix $\mathbf{M}(s,T)$ are $M_{11}=-\gamma_{23}/2+\alpha(T)/(sL)$, $M_{12}=i\Omega_d/2$, $M_{21}=i\Omega_d^*/2+\beta(T)/(sL)$, and $M_{22}=-\gamma_{21}/2$. Here, $\alpha(T)=\frac{|g_s|^2NL}{c}\bigl[\langle\hat{\sigma}_{33}(T)\rangle-\langle\hat{\sigma}_{22}(T)\rangle\bigr]$ governs the gain or absorption and dispersion of the propagating Stokes field, while $\beta(T)=\frac{|g_s|^2NL}{c}\langle\hat{\sigma}_{31}(T)\rangle$ describes its coupling to the spin-wave coherence through the write-induced optical coherence. The inhomogeneous source vector is $\mathbf{N}(s,T)=-\frac{ic}{g_s^*N}\frac{1}{sL}\begin{bmatrix}\alpha(T),\beta(T)\end{bmatrix}^{\rm T}\hat{a}_s(0,T)+\begin{bmatrix}\hat{G}_{23}(s,T),\hat{G}_{21}(s,T)\end{bmatrix}^{\rm T}\equiv\mathbf{N}_{\rm vac}(s,T)+\mathbf{N}_{\rm L}(s,T)$, where $\mathbf{N}_{\rm vac}$ represents the Stokes vacuum boundary input and $\mathbf{N}_{\rm L}$ contains the Langevin noise operators.

During the write process, the atomic populations and optical coherence evolve, making $\alpha(T)$ and $\beta(T)$ time dependent. Consequently, $\mathbf{M}(s,T)$ generally does not commute at different times, and the evolution requires a time-ordered propagator. We evaluate it using a piecewise-constant construction. For a discrete time step $\Delta T=T_{k+1}-T_k$, we treat the coefficient matrix as constant within the interval $[T_k,T_{k+1}]$. The source vector retains its time dependence within each interval. The corresponding local solution is
\begin{align}
	\mathbf{S}(s, T_{k+1})
	=&
	e^{\mathbf{M}(s, T_k) \Delta T} \mathbf{S}(s, T_k)
	\nonumber\\&+
	\int_{T_k}^{T_{k+1}} e^{\mathbf{M}(s, T_k)(T_{k+1} - T')} \mathbf{N}(s, T') dT'
	.\label{eq10}
\end{align}
In the limit $\Delta T\rightarrow0$, the resulting piecewise construction converges to the time-ordered evolution generated by $\mathbf{M}(s,T)$. Defining $U_k=\exp[\mathbf{M}(s,T_k)\Delta T]$ and $\mathbf{W}(T_n,T_{k+1})=\prod_{j=k+1}^{n-1}U_j$, iteration of the local propagators gives
\begin{widetext}
	\begin{align}
		&\mathbf{S}(s, T_n)
		=
		\left( \prod_{k=0}^{n-1} U_k \right) \mathbf{S}(s, 0)
		+ \sum_{k=0}^{n-1} \left( \prod_{j=k+1}^{n-1} U_j \right) \int_{T_k}^{T_{k+1}} e^{\mathbf{M}(s, T_k)(T_{k+1} - T')} \mathbf{N}(s, T') dT'
		\nonumber\\
		&=
		\mathbf{W}(T_n, 0) \mathbf{S}(s, 0)
		+ \sum_{k=0}^{n-1} \int_{T_k}^{T_{k+1}} \left[ \mathbf{W}(T_n, T_{k+1}) e^{\mathbf{M}(s, T_k)(T_{k+1} - T')} \right] \mathbf{N}(s, T') dT'
		=
		\mathbf{W}(T_n, 0) \mathbf{S}(s, 0) + \int_{0}^{T_n} \mathbf{w}(T_n, T') \mathbf{N}(s, T') dT'
		.\label{eq11}
	\end{align}
\end{widetext}
Here and throughout the following derivation, the products are time ordered with later-time operators acting from the left, so that $\prod_{k=0}^{n-1}U_k\equiv U_{n-1}\cdots U_1U_0$. The transfer matrix $\mathbf{w}(T_n,T')\equiv\mathbf{W}(T_n,T_{k+1})e^{\mathbf{M}(s,T_k)(T_{k+1}-T')}$ propagates a source introduced at $T'\in[T_k,T_{k+1}]$ to the final time $T_n$. In the continuous-time limit, we replace $T_n$ by an arbitrary observation time $T$. Writing $\mathbf{W}=[W_{ij}]$ and $\mathbf{w}=[w_{ij}]$, the solution becomes
\begin{widetext}
	\begin{align}
		&\begin{bmatrix} \hat{S}_{23}(s, T) \\ \hat{S}_{21}(s, T) \end{bmatrix}
		=
		\begin{bmatrix}
			W_{11}(T, 0) & W_{12}(T, 0) \\
			W_{21}(T, 0) & W_{22}(T, 0)
		\end{bmatrix}
		\begin{bmatrix}
			\hat{S}_{23}(s, 0) \\ \hat{S}_{21}(s, 0)
		\end{bmatrix}
		\nonumber\\
		&-
		\frac{ic}{g_{s}^{*} N sL} \int_{0}^{T}
		\begin{bmatrix}
			w_{11}(T, T') & w_{12}(T, T') \\
			w_{21}(T, T') & w_{22}(T, T')
		\end{bmatrix}
		\begin{bmatrix}
			\alpha(T') \\ \beta(T')
		\end{bmatrix}
		\hat{a}_{s}(0, T') dT'
		+
		\int_{0}^{T}
		\begin{bmatrix}
			w_{11}(T, T') & w_{12}(T, T') \\
			w_{21}(T, T') & w_{22}(T, T')
		\end{bmatrix}
		\begin{bmatrix}
			\hat{G}_{23}(s, T') \\ \hat{G}_{21}(s, T')
		\end{bmatrix} dT'. \label{eq12}
	\end{align}
\end{widetext}
Here, $W_{ij}$ describe the homogeneous evolution of the initial collective atomic operators, whereas $w_{ij}$ propagate the Stokes vacuum boundary input and the Langevin noise generated during the write process. The corresponding real-space coherences are obtained by inverse Laplace transformation.

To calculate the Stokes photon statistics and determine the spin-wave initial condition for the subsequent retrieval stage, we express these solutions in real space and time. After inverse transformation and integration of the field equation, the Stokes field and spin-wave operators are
\begin{widetext}
	\begin{align}
		\hat{a}_s(Z,T)
		=&
		\hat{a}_s(0,T)+
		\int_0^T dT'
		\mathcal{G}(Z,T,T')\hat{a}_s(0,T')
		+
		\frac{ig_s^*N}{c}\int_0^Z dZ'
		\begin{bmatrix}
			\mathcal{G}^{W}_{11}(Z',T) \\ \mathcal{G}^{W}_{12}(Z',T)
		\end{bmatrix}^{\rm T}
		\begin{bmatrix}
			\hat{\sigma}_{23}(Z-Z',0) \\ \hat{\sigma}_{21}(Z-Z',0)
		\end{bmatrix}
		\nonumber\\
		&+
		\frac{ig_s^*N}{c}\int_0^T dT'\int_0^Z dZ'
		\begin{bmatrix}
			\mathcal{G}^{w}_{11}(Z',T,T') \\ \mathcal{G}^{w}_{12}(Z',T,T')
		\end{bmatrix}^{\rm T}
		\begin{bmatrix}
			\hat{F}_{23}(Z-Z',T') \\ \hat{F}_{21}(Z-Z',T')
		\end{bmatrix}
		,\label{eq13} \\
		\hat{\sigma}_{21}(Z,T)
		=&
		-\frac{ic}{g_s^*NL}\int_0^T dT'
		\mathcal{G}_{\rm spin}(Z,T,T')\hat{a}_s(0,T')
		+
		\int_0^Z dZ'
		\begin{bmatrix}
			\mathcal{G}^{W}_{21}(Z',T) \\ \mathcal{G}^{W}_{22}(Z',T)
		\end{bmatrix}^{\rm T}
		\begin{bmatrix}
			\hat{\sigma}_{23}(Z-Z',0) \\ \hat{\sigma}_{21}(Z-Z',0)
		\end{bmatrix}
		\nonumber\\
		&+
		\int_0^T dT'\int_0^Z dZ'
		\begin{bmatrix}
			\mathcal{G}^{w}_{21}(Z',T,T') \\ \mathcal{G}^{w}_{22}(Z',T,T')
		\end{bmatrix}^{\rm T}
		\begin{bmatrix}
			\hat{F}_{23}(Z-Z',T') \\ \hat{F}_{21}(Z-Z',T')
		\end{bmatrix}
		.\label{eq14}
	\end{align}
\end{widetext}
Here, $\mathcal{G}(Z,T,T')=\mathcal{L}^{-1}[s^{-2}(w_{11}\alpha+w_{12}\beta)]/L$ and $\mathcal{G}_{\rm spin}(Z,T,T')=\mathcal{L}^{-1}[s^{-1}(w_{21}\alpha+w_{22}\beta)]$ describe the Stokes-field and spin-wave responses to the boundary vacuum input. The functions $\mathcal{G}^{W}_{1j}(Z,T)=\mathcal{L}^{-1}[s^{-1}W_{1j}]$ and $\mathcal{G}^{W}_{2j}(Z,T)=\mathcal{L}^{-1}[W_{2j}]$, with $j\in\{1,2\}$, describe the contributions from the initial atomic operators. The functions $\mathcal{G}^{w}_{1j}(Z,T,T')=\mathcal{L}^{-1}[s^{-1}w_{1j}]$ and $\mathcal{G}^{w}_{2j}(Z,T,T')=\mathcal{L}^{-1}[w_{2j}]$ describe the corresponding Langevin-noise contributions. All inverse Laplace transforms are understood as causal one-way spatial response kernels on the positive half-axis. Equations~\eqref{eq13} and \eqref{eq14} therefore distinguish the propagated boundary vacuum input, the initial collective atomic fluctuations, and the Langevin noise generated and accumulated during the write process.

The Stokes photon generation rate is evaluated from the output flux relation $R_s(T)=\frac{c}{L}\langle\hat{a}_s^\dagger(L,T)\hat{a}_s(L,T)\rangle$. This calculation incorporates the relevant local atomic correlations of the form $\langle\hat{\sigma}_{jk}(Z_1,T)\hat{\sigma}_{j'k'}(Z_2,T)\rangle=\frac{L}{N}\delta_{j'k}\langle\hat{\sigma}_{jk'}(Z_1,T)\rangle\delta(Z_1-Z_2)$, together with the Langevin-noise correlations. Under the Markovian approximation, $\langle\hat{F}_{jk}(Z_1,T_1)\hat{F}_{j'k'}(Z_2,T_2)\rangle=\frac{L}{N}\mathcal{D}_{jk,j'k'}(T_1)\delta(Z_1-Z_2)\delta(T_1-T_2)$, where $\mathcal{D}_{jk,j'k'}$ is the diffusion coefficient obtained from the generalized Einstein relation. Detailed derivations are provided in \setcitestyle{numbers}Refs.~\cite{JS1,JS3}.\setcitestyle{super}

To characterize the Stokes photon statistics, we evaluate the zero-delay second-order autocorrelation function, $g^{(2)}_{s\text{-}s}(0)=\langle\hat{a}_{s}^{\dagger}\hat{a}_{s}^{\dagger}\hat{a}_{s}\hat{a}_{s}\rangle/\langle\hat{a}_{s}^{\dagger}\hat{a}_{s}\rangle^{2}$. Since $\hat{a}_{s}(L,T)$ is a linear combination of the boundary vacuum input, the initial atomic fluctuations, and the Langevin sources, its fourth-order moment can be evaluated using Wick's theorem under the Gaussian-noise approximation. For the zero-mean, phase-insensitive Stokes field considered here, $\langle\hat{a}_s^\dagger\hat{a}_s^\dagger\hat{a}_s\hat{a}_s\rangle=2\langle\hat{a}_s^\dagger\hat{a}_s\rangle^2$, giving $g^{(2)}_{s\text{-}s}(0)=2$. This result is the thermal limit for the time-resolved Stokes field in the selected spatial mode. Integration over a finite detection window can include multiple independent temporal modes and may therefore yield a measured value below two.


\subsection*{Large-detuning regime}

We first examine the large-detuning regime, which is widely used in DLCZ-type implementations and provides a transparent limiting case for benchmarking the open-system theory. To distinguish detuning effects from population depletion, we begin by deriving an analytic response in the strict undepleted limit and then introduce a phenomenological correction for the leading depletion effect.

When $|\Delta_d|\gg\Gamma_3$ and $|\Delta_d|\gg|\Omega_d|$, the instantaneous excited-state population remains small. The additional approximation $\langle\hat{\sigma}_{11}\rangle\simeq1$ requires weak accumulated optical pumping, since large detuning alone does not prevent appreciable population transfer during an extended write pulse. In the undepleted limit, the mean optical coherence $\langle\hat{\sigma}_{31}\rangle$ associated with the $|1\rangle\leftrightarrow|3\rangle$ transition follows the driving field on a fast time scale. Adiabatic elimination of this coherence in Eq.~\eqref{eq5} gives $\langle\hat{\sigma}_{31}\rangle\simeq-i\Omega_d^{*}/(\Gamma_3+2i\Delta_d)$. Writing the total spontaneous decay rate as
$\Gamma_3\equiv\Gamma$, with $\Gamma=2\pi\times 6$~MHz for the $^{87}$Rb D$_2$ transition, and denoting the resonant optical depth by OD, the Stokes photon generation rate in the strict undepleted large-detuning limit reduces to
\begin{align}
	R_s =& R_{s1}+R_{s2},
	\qquad
	R_{s1}=\frac{\Gamma\mathrm{OD}}{4L}\int_0^L dZ\,|\mathcal{G}^W_{12}(Z,T)|^2.
	\label{eq15}
\end{align}
Here, $R_{s1}$ represents the homogeneous vacuum-seeded contribution associated with the initial atomic fluctuations. The second term $R_{s2}$ accounts for the Langevin-noise contribution and is evaluated as $R_{s2}=\frac{\Gamma\mathrm{OD}}{4L}\int_0^L dZ\int_0^T dT'\,\mathbf{v}^\dagger(Z,T-T')\mathbf{D}(T')\mathbf{v}(Z,T-T')$. The response vector is defined as $\mathbf{v}(Z,T-T')=[\mathcal{G}^W_{11}(Z,T-T'),\mathcal{G}^W_{12}(Z,T-T')]^{\rm T}$, and the noise diffusion matrix is given by $\mathbf{D}(T')=\left[\begin{smallmatrix}\mathcal{D}_{32,23}(T')&\mathcal{D}_{32,21}(T')\\ \mathcal{D}_{12,23}(T')&\mathcal{D}_{12,21}(T')\end{smallmatrix}\right]$, whose elements are $\mathcal{D}_{32,23}(T')=0$, $\mathcal{D}_{32,21}(T')=\tfrac{\gamma_{21}}{2}\langle\hat{\sigma}_{31}(T')\rangle=\mathcal{D}_{12,23}^*(T')$, and $\mathcal{D}_{12,21}(T')=\gamma_{21}\langle\hat{\sigma}_{11}(T')\rangle+\tfrac{\Gamma}{2}\langle\hat{\sigma}_{33}(T')\rangle$. In the strict undepleted limit, these coefficients are evaluated using $\langle\hat{\sigma}_{11}\rangle=1$, $\langle\hat{\sigma}_{33}\rangle=0$, and $\langle\hat{\sigma}_{31}\rangle=-i\Omega_d^{*}/(\Gamma+2i\Delta_d)$.

The Stokes coupling strength is related to OD through $|g_s|^2NL/c=\Gamma\mathrm{OD}/4$. For the level configuration considered here, we take $\Gamma_{31}=\Gamma_{32}=\Gamma/2$. These total branching rates are distinct from the coupling to the selected phase-matched Stokes mode represented by $g_s$. Equation~\eqref{eq15}, together with the analytic Green's functions given in Methods, defines the strict undepleted large-detuning model. To estimate the leading depletion effect, we replace $\mathrm{OD}$ by the time-dependent effective optical depth $\mathrm{OD}_{\mathrm{eff}}(T)=\mathrm{OD}\langle\hat{\sigma}_{11}(T)\rangle$ in the analytic response functions entering both $R_{s1}$ and $R_{s2}$. For the Langevin contribution $R_{s2}$, the diffusion matrix $\mathbf{D}(T')$ is correspondingly evaluated using the time-dependent mean atomic values rather than their constant undepleted-limit values. This quasistatic correction captures the leading effects of population redistribution on the Raman response and Langevin-noise contribution, but not the complete time-dependent atomic, noise, and propagation dynamics. It is therefore used only as an intermediate comparison with the full open-system calculation. The parameters used in this regime are summarized in Methods.

\FigTwo

For comparison, we also consider the foundational undepleted-medium model of \setcitestyle{numbers}Ref.~\cite{Raymer},\setcitestyle{super} which describes the evolution from spontaneous Raman initiation to stimulated buildup. This model assumes undepleted ground-state populations and does not retain the transient optical response or the explicit propagation of Langevin noise included in the present treatment. The resulting rate equation is
\begin{align}
	R_s = \Gamma_{\rm R}B(T)+\gamma_{21}\Gamma_{\rm R}\int_0^T dT'B(T')
	,\label{eq16}
\end{align}
where $\Gamma_{\rm R}=\Gamma\mathrm{OD}|\Omega_d|^2/(16\Delta_d^2)$ sets the characteristic Stokes generation rate determined by the effective Raman coupling. The function $B(T)=e^{-\gamma_{21}T}[I_0^2(\sqrt{4\Gamma_{\rm R}T})-I_1^2(\sqrt{4\Gamma_{\rm R}T})]$ describes the temporal buildup of SRS within this approximation, where $I_0$ and $I_1$ are modified Bessel functions of the first kind.

Figure~\ref{fig2} compares the Stokes photon generation dynamics predicted by the different theoretical descriptions, with the temporal origin defined by the onset of the driving field. Figure~\ref{fig2}(a) shows that the open-system theory captures both the early-time oscillations, arising from transient coherence between $|1\rangle$ and $|3\rangle$ with a period of approximately $11.1$~ns, and the subsequent transient gain followed by a gradual decrease. The decrease results from population transfer to $|2\rangle$, which reduces the population available in $|1\rangle$ for further Raman excitation.

The strict undepleted large-detuning response applies only before appreciable optical pumping develops. The depletion-corrected estimate shown by the blue solid curve reproduces the initial dynamics and approximately captures the subsequent gain suppression through the reduction of $\mathrm{OD}_{\mathrm{eff}}(T)$. Its deviation from the full theory at longer times reflects the limitations of the quasistatic correction. The steady-state SFWM rate shown by the black dotted curve is close to the early-time average of the DLCZ dynamics before appreciable spin-wave accumulation. The subsequent enhancement is absent in continuous SFWM because the spin-wave coherence is continuously converted into anti-Stokes photons, limiting the collective buildup during the write process.

By contrast, the undepleted-medium model~\cite{Raymer} shown by the red dashed curve reproduces the correct early-time order of magnitude within its intended regime but does not include the transient optical response or population redistribution. Although it describes stimulated Raman buildup, the generation rate continues to increase and eventually approaches a plateau without the characteristic decrease caused by depletion. Its growing deviation for extended write pulses therefore reflects the breakdown of the undepleted-population approximation.

Figures~\ref{fig2}(b)--\ref{fig2}(d) show how the Stokes generation dynamics depend on the driving parameters and OD. Reducing the driving strength $\Omega_d$ [Fig.~\ref{fig2}(b)] or increasing the detuning $\Delta_d$ [Fig.~\ref{fig2}(c)] suppresses the early-time rate according to $R_s\propto|\Omega_d/\Delta_d|^2$ and slows the population-transfer dynamics. Increasing the OD [Fig.~\ref{fig2}(d)] enhances the early-time rate approximately linearly, consistent with $R_s\propto\mathrm{OD}|\Omega_d/\Delta_d|^2$, while preserving the qualitative transient buildup and depletion behavior. Before substantial spin-wave accumulation develops, the time-averaged DLCZ rate remains close to the steady-state SFWM prediction under identical driving conditions. At later times, spin-wave accumulation during the DLCZ write process produces the pronounced departure from the continuous SFWM response.


\subsection*{Storage and readout dynamics}

The write--store--read sequence in the DLCZ protocol establishes a controllable mapping between collective atomic coherence and propagating optical modes. During the write process, Stokes-photon generation is accompanied by the creation of the spin-wave coherence $\hat{\sigma}_{21}$ across the atomic ensemble. After the driving pulse is terminated, the stored coherence evolves freely and, for $T>T_d$, decays as
\begin{equation}
	\hat{\sigma}_{21}(Z,T) = \hat{\sigma}_{21}(Z,T_d)\, e^{-\frac{\gamma_{21}}{2}(T - T_d)}
	,\label{eq17}
\end{equation}
where $T_d$ denotes the end time of the driving pulse. The storage-stage Langevin term is omitted here because, for the vacuum reservoir considered, it does not add excitations to the normally ordered retrieval observables. Within the spin-wave lifetime, a coupling field resonant with the $|2\rangle \leftrightarrow |4\rangle$ transition initiates the readout process and couples the stored excitation to the anti-Stokes mode.

During retrieval, the coupling field coherently maps the stored spin-wave excitation onto the anti-Stokes field $\hat{a}_{as}$. The interaction Hamiltonian governing this process is
\begin{equation}
	\hat{H}_{\rm R}
	=-\frac{\hbar N}{2L}\int_0^L dz
	\left(
	\Delta_c \hat{\sigma}_{44}
	+\Omega_c \hat{\sigma}_{42}
	+2 g_{as} \hat{a}_{as} \hat{\sigma}_{41}
	+\mathrm{h.c.}
	\right),
\end{equation}
where $\Omega_c$ and $\Delta_c$ are the Rabi frequency and detuning of the coupling field, respectively. The state $|4\rangle$ denotes the excited-state sublevel addressed by the read pathway in the effective level scheme. In the counter-propagating geometry, where the coupling field $\Omega_c$ propagates opposite to the driving field $\Omega_d$, phase matching selects the backward-propagating anti-Stokes mode. The retrieval dynamics are described by the MSE coupled to the corresponding HLEs,
\begin{align}
	&\left(\frac{1}{c}\frac{\partial}{\partial t}-\frac{\partial}{\partial z}\right)\hat{a}_{as}^\dagger=-\frac{ig_{as}N}{c}\hat{\sigma}_{41}
	-i\Delta k\hat{a}_{as}^\dagger
	,\label{eq19}
	\\
	&\frac{\partial\hat{\sigma}_{41}}{\partial t}=
	-\frac{i}{2}\Omega_c^*\hat{\sigma}_{21}
	-ig_{as}^*\hat{a}_{as}^\dagger
	-\frac{\gamma_{41}}{2}\hat{\sigma}_{41}+\hat{F}_{41}
	,\label{eq20}
	\\
	&\frac{\partial\hat{\sigma}_{21}}{\partial t}=
	-i\frac{\Omega_c}{2}\hat{\sigma}_{41}
	-\frac{\gamma_{21}}{2}\hat{\sigma}_{21}+\hat{F}_{21}
	,\label{eq21}
\end{align}
where $\gamma_{41}=\Gamma_4+2i\Delta_c$ is the complex optical-coherence decay parameter. Equations~\eqref{eq19}--\eqref{eq21} describe retrieval in the linear weak-excitation regime, where the atomic population remains predominantly in $|1\rangle$ and read-induced changes of the mean populations are negligible. The model therefore describes coherent retrieval from the phase-matched spin wave but does not include fluorescence associated with incoherent population transferred to $|2\rangle$. Here, $\Delta k$ denotes the longitudinal wave-vector mismatch for phase-matched retrieval into the backward anti-Stokes mode; we take $\Delta kL=0.37\pi$ for the configuration considered here. The backward-propagating anti-Stokes input at $z=L$ is taken to be vacuum, while the initial spin-wave operator at the onset of retrieval is given by Eq.~\eqref{eq17} after the chosen storage interval.

\FigThree

To describe the backward-propagating geometry, we introduce the moving-frame coordinates $\bar{Z}=z$ and $\bar{T}=t+z/c$. Under this transformation, the derivatives become $\partial_{\bar{Z}}=\partial_z-c^{-1}\partial_t$ and $\partial_{\bar{T}}=\partial_t$, reducing the propagation equation for the backward anti-Stokes mode to a spatial evolution equation. After obtaining the field operator $\hat{a}_{as}(\bar{Z},\bar{T})$, we evaluate the retrieved anti-Stokes photon rate at the backward output boundary from the flux relation $R_{as}(\bar{T})=\tfrac{c}{L}\langle\hat{a}_{as}^\dagger(0,\bar{T})\hat{a}_{as}(0,\bar{T})\rangle$.

Figure~\ref{fig3}(a) shows the tunable retrieval dynamics of the anti-Stokes field. To remain within the weak-excitation regime considered here, the calculation uses a 100-ns driving pulse followed by a 200-ns storage interval, with $\Gamma_4=\Gamma_3=\Gamma$. In the weaker-coupling regime, $\Lambda$-type electromagnetically induced transparency (EIT)~\cite{EIT1,EIT2,EIT3} produces a delayed and temporally broadened anti-Stokes wavepacket. Increasing the coupling strength broadens the transparency window and produces faster and more temporally compressed retrieval. The coupling field therefore provides direct control over the temporal mode of the retrieved photon.

The coupling detuning $\Delta_c$ provides an additional control parameter, as shown in Fig.~\ref{fig3}(b). Near resonance, increasing $\Delta_c$ from $0$ to $1\Gamma$ suppresses the EIT slow-light response and reduces the retrieval delay. At larger detuning, the weaker effective atom--field interaction increases the retrieval timescale, as illustrated by $\Delta_c=3\Gamma$. The retrieval delay therefore varies nonmonotonically with detuning. The asymmetry between $\Delta_c=\pm3\Gamma$ arises from the interplay between the finite phase mismatch in the backward geometry and the detuning-dependent spectral response, consistent with the asymmetric behavior found in SFWM~\cite{JS2}.

\FigFour

The retrieved anti-Stokes waveforms exhibit distinct dynamical regimes. In the low-OD regime with strong coupling, the waveform shows Rabi-like oscillations [Fig.~\ref{fig4}(a)] arising from coherent exchange between the stored spin-wave excitation and the excited-state optical coherence mediated by the coupling field. The damping is governed primarily by single-atom spontaneous decay, so increasing $\Omega_c$ shortens the oscillation period without substantially changing the overall retrieval envelope. By contrast, in the high-OD regime with strong coupling, collective emission shortens the retrieval timescale beyond that set by single-atom decay and produces superradiant retrieval~\cite{super1,super2,super3} [Fig.~\ref{fig4}(b)]. Constructive interference among the phase-matched atomic emission amplitudes releases the stored excitation as a pulse shorter than the single-atom lifetime.

When the coupling field is reduced in the high-OD regime, a sharp transient front appears before the main slow-light wavepacket, forming an optical precursor [Fig.~\ref{fig4}(c)]. This early-time component originates from rapidly varying spectral components generated at the onset of the coupling field, which propagate differently from the main wavepacket in the dispersive medium. Its temporal advance results from waveform reshaping and does not represent superluminal information transfer.


\subsection*{Two-photon correlation properties}

To quantify the joint temporal correlations between the generated Stokes and anti-Stokes photons, we evaluate the normalized second-order cross-correlation function
\begin{align}
	g_{s\text{-}as}^{(2)}(T,\bar{T})=\frac
	{\langle\hat{a}_s^\dagger(T)\hat{a}_{as}^\dagger(\bar{T})\hat{a}_{as}(\bar{T})\hat{a}_s(T)\rangle}
	{\langle\hat{a}_s^\dagger(T)\hat{a}_s(T)\rangle\langle\hat{a}_{as}^\dagger(\bar{T})\hat{a}_{as}(\bar{T})\rangle},
	\label{eq22}
\end{align}
where $\hat{a}_s(T)\equiv\hat{a}_s(L,T)$ and $\hat{a}_{as}(\bar{T})\equiv\hat{a}_{as}(0,\bar{T})$ are the output field operators. For statistically independent fields, $g_{s\text{-}as}^{(2)}=1$, whereas values exceeding unity indicate positive temporal correlations between the Stokes and anti-Stokes fields. Equation~\eqref{eq22} describes the time-resolved correlation between photons emitted at the specific times $T$ and $\bar{T}$. The experimentally measured integrated correlation introduced below is instead a photon-flux-weighted average over finite Stokes and anti-Stokes detection windows. The peak and integrated correlations therefore represent distinct temporal observables and are not generally identical.

Figure~\ref{fig5}(a) shows the calculated cross-correlation function, where $T$ and $\bar{T}$ denote the Stokes-generation and anti-Stokes-retrieval times, respectively. Compared with the Stokes temporal profile, the oscillatory modulation along the correlation ridge is strongly suppressed. For a fixed write time $T$, the conditional anti-Stokes temporal profile encoded in the cross-correlation differs markedly from the unconditional anti-Stokes intensity profile shown in Fig.~\ref{fig3}. The joint temporal statistics therefore cannot be inferred from the single-channel intensities alone.

\FigFive

The effect of increasing the coupling strength $\Omega_c$ is shown in Fig.~\ref{fig5}(b). A stronger coupling field accelerates the anti-Stokes retrieval and compresses the correlation profile along the anti-Stokes time axis. Despite this compression, the peak cross-correlation remains nearly unchanged over the retrieval conditions considered. This behavior differs from the corresponding SFWM calculation, in which increasing $\Omega_c$ reduces the slow-light delay and increases the correlation peak under otherwise identical conditions.

This distinction follows from the different scaling of the conditional coincidence signal and accidental background. In SFWM, reducing $\Omega_c$ narrows the anti-Stokes bandwidth and temporally broadens the wavepacket. The resulting temporal stretching lowers the peak conditional coincidence rate, while the unconditioned anti-Stokes rate and accidental background vary more weakly. In DLCZ-type SRS, the read pulse acts on a spin wave created before retrieval. Changes in the retrieval parameters therefore reshape the conditional readout signal and accidental background approximately in parallel, leaving the peak cross-correlation nearly unchanged over the range considered. A similar response is obtained when varying the coupling detuning, as shown in Fig.~\ref{fig5}(c).

The classical control over the retrieval timing highlights a practical feature of DLCZ-type SRS relevant to temporal-mode encoding and high-dimensional quantum information processing~\cite{tbin1,tbin2}. In some heralded SFWM~\cite{tbin3} and spontaneous parametric down-conversion~\cite{tbin4} schemes, temporal processing relies on herald-conditioned modulation, switching, or post-selection. In DLCZ-type SRS, the classical read pulse predetermines the retrieval window and enables temporal selection, slicing, or gating of the anti-Stokes wavepacket without using Stokes-photon detection as a real-time feed-forward trigger.

The full time-resolved profile of $g_{s\text{-}as}^{(2)}$ can also be used to evaluate the pairing ratios, defined as $r_{p,l}=N_{\rm corr}/N_l$ for $l\in\{s,as\}$. Here, $N_{\rm corr}=\int_0^{T_d}dT\int_0^\infty d\bar{T}\,R_s(T)R_{as}(\bar{T})[g_{s\text{-}as}^{(2)}(T,\bar{T})-1]$ represents the number of correlated photon pairs, while $N_s=\int_0^{T_d}dT\,R_s(T)$ and $N_{as}=\int_0^\infty d\bar{T}\,R_{as}(\bar{T})$ denote the total numbers of generated Stokes and anti-Stokes photons, respectively. For negligible spin-wave decoherence, the pairing ratios are only weakly affected by the retrieval conditions considered in Fig.~\ref{fig5} and are governed primarily by the collective OD. For the retrieval conditions shown in Fig.~\ref{fig5}, the calculated pairing ratios remain approximately $r_{p,s}=0.61$ and $r_{p,as}=0.70$, comparable to the SFWM results. When the OD is increased to 100, both $r_{p,s}$ and $r_{p,as}$ approach approximately $0.9$ owing to strong collective enhancement~\cite{JS1}.

Beyond the two-photon cross-correlation, the single-photon character of the heralded anti-Stokes field can be characterized by the zero-delay conditional autocorrelation function~\cite{JS3}. Figure~\ref{fig5}(d) shows the calculated conditional autocorrelation profile under the retrieval conditions of Fig.~\ref{fig5}(a). Within the zero-mean Gaussian statistical model used here, the time-resolved conditional anti-Stokes autocorrelation is inferred from the cross-correlation as $g_{as\text{-}as|s}^{(2)}(T,\bar{T})=[4g_{s\text{-}as}^{(2)}(T,\bar{T})-2][g_{s\text{-}as}^{(2)}(T,\bar{T})]^{-2}$. For a strongly correlated Stokes--anti-Stokes pair, the conditional autocorrelation is suppressed below unity, indicating nonclassical suppression of multiphoton contributions in the heralded anti-Stokes field. This model-derived quantity is time resolved and is not directly equivalent to a conditional Hanbury Brown--Twiss measurement integrated over finite detection windows. As $g_{s\text{-}as}^{(2)}$ approaches the uncorrelated limit of unity, $g_{as\text{-}as|s}^{(2)}$ approaches 2, corresponding to the thermal statistics of the unconditioned anti-Stokes field within the model.

\FigSix


\subsection*{Experimental observations}

For quantitative comparison with experiment, the calculations use the hyperfine-summed branching fractions introduced above, while polarization selection of the collected phase-matched Stokes and anti-Stokes modes is incorporated through the effective atom--field coupling constants.

Figure~\ref{fig6}(a) shows the experimentally measured integrated cross-correlation, defined as $\bar{g}_{s\text{-}as}^{(2)}=p_{s,as}/(p_s p_{as})$. Here, $p_s$ and $p_{as}$ denote the unconditional single-detection probabilities in the Stokes and anti-Stokes channels within the windows $\Delta T$ and $\Delta\bar{T}$, respectively, while $p_{s,as}$ is the coincidence probability. The data were acquired with a 200-ns storage time. The Stokes window is fixed at $\Delta T=100$~ns, whereas the anti-Stokes window $\Delta\bar{T}$ is adjusted for each parameter set to include the temporal profile until its intensity falls to $1/e$ of its peak. The measured $\bar{g}_{s\text{-}as}^{(2)}$ depends only weakly on $\Omega_c$, in agreement with the theoretical prediction 
\begin{align}
	\bar{g}_{s\text{-}as}^{(2)}
	=
	\int_{T_0}^{T_0+\Delta T} \! dT
	\int_{\bar{T}_0}^{\bar{T}_0+\Delta\bar{T}} \! d\bar{T}
	\,
	f(T,\bar{T}) g_{s\text{-}as}^{(2)}(T,\bar{T}),
	\label{eq23}
\end{align}
where $f(T,\bar{T})$ is a normalized weighting function proportional to $R_s(T)R_{as}(\bar{T})$ within the same integration windows. Using the experimental parameters and integration windows, with leakage light and detector dark counts also taken into account, the model yields $\bar{g}_{s\text{-}as}^{(2)}\approx15$. The measured values are consistent with this estimate in overall magnitude.

\FigSeven

Figure~\ref{fig6}(b) compares the time-resolved peak $g_{s\text{-}as}^{(2)}$ values predicted for DLCZ-type SRS and SFWM. These peak values characterize the temporal correlation response and are distinct from the finite-window measurements in Figs.~\ref{fig6}(a) and \ref{fig6}(c). Quantitative comparison with the measured integrated correlations is made using Eq.~\eqref{eq23}. The SFWM curves are based on the open-system model validated in our previous studies~\cite{JS1,JS3}. In the DLCZ scheme, the weak dependence on $\Omega_c$ indicates that the retrieved-photon bandwidth can be reduced without substantially lowering the Stokes--anti-Stokes correlation, providing flexibility for bandwidth engineering~\cite{bandwidth}. Measurements at different coupling detunings $\Delta_c$, shown in Fig.~\ref{fig6}(c), demonstrate the frequency tunability of the heralded anti-Stokes photon. The peak calculations in Fig.~\ref{fig6}(d) predict only a weak change in the DLCZ correlation over the range considered, in contrast to the stronger reduction for SFWM. This spectral flexibility is relevant for interfacing the source with heterogeneous quantum nodes, including quantum memories~\cite{QM1,QM2,QM3,QM4} and quantum frequency conversion modules~\cite{QFC1,QFC2,QFC3,QFC4,QFC5,QFC6,QFC7}, where spectral matching is required.

Figures~\ref{fig7}(a)--\ref{fig7}(c) show the integrated cross-correlation as a function of retrieval time $T_{\rm ret}$ for different driving detunings $\Delta_d$. Reducing $\Delta_d$ increases the Raman excitation probability and Stokes generation rate, leading to a larger mean spin-wave excitation and a greater relative accidental-coincidence contribution. Increasing $\Delta_d$ has the opposite effect and enhances $\bar{g}_{s\text{-}as}^{(2)}$, consistent with the large-detuning scaling $R_s\propto|\Omega_d/\Delta_d|^2$. The theoretical comparison in Fig.~\ref{fig7}(d) shows that both DLCZ-type SRS and SFWM exhibit an approximate $\Delta_d^2$ scaling of the peak $g_{s\text{-}as}^{(2)}$ over the detuning range considered. Figure~\ref{fig7}(d) also compares DLCZ-type SRS for two write-pulse durations, $T_d=100$~ns and 20~ns. The two DLCZ curves further reveal a strong dependence on the write-pulse duration. Reducing the pulse width from 100~ns to 20~ns substantially enhances the predicted correlation, with the 20-ns DLCZ result exceeding the corresponding SFWM prediction under the same modeled conditions. This enhancement arises because the correlated coincidence signal and accidental background scale differently with the mean spin-wave excitation as the write-pulse duration is varied.

Spin-wave decoherence during storage also affects the measured correlation. In the present system, residual dephasing from magnetic-field inhomogeneity and motional dephasing due to the nonzero angle between the coupling and anti-Stokes modes~\cite{angle1,angle2,angle3} reduce the correlation contrast with increasing storage time. Together, these mechanisms gradually reduce the stored collective spin-wave coherence. This behavior is consistent with an approximately exponential decrease of $\bar{g}_{s\text{-}as}^{(2)}-1$ at longer storage times.

\FigEight

Figure~\ref{fig8} clarifies the role of the write-pulse duration $T_d$ in determining the two-photon correlation. The measured $\bar{g}_{s\text{-}as}^{(2)}$ progressively decreases as $T_d$ increases, confirming that shorter write pulses enhance the Stokes--anti-Stokes correlation. A longer write pulse creates a larger mean number of Stokes photons and spin-wave excitations, causing the accidental-coincidence contribution to increase more rapidly relative to the correlated signal. The theoretical curve in Fig.~\ref{fig8} includes a $T_d$-independent background contribution from leakage light and detector dark counts. Over the pulse-duration range considered, these backgrounds account for approximately 5\% of the detected counts in each of the Stokes and anti-Stokes channels and consequently reduce the measured cross-correlation.

The theoretical model captures this trend and further predicts $\bar{g}_{s\text{-}as}^{(2)}\approx88$ at $T_d=10$~ns, below the experimentally accessible pulse-duration range. This value substantially exceeds that obtained with a 100-ns write pulse and the corresponding SFWM prediction under comparable conditions. Shortening the write pulse therefore provides an effective route to correlation enhancement and is expected to improve the single-photon character of the heralded anti-Stokes field by reducing the relative multiphoton and accidental contributions.

\FigNine

Beyond shaping the Stokes generation process, the DLCZ protocol also supports temporal selection of the retrieved anti-Stokes wavepacket. Figure~\ref{fig9} presents an experimental temporal-slicing analysis of the Stokes--anti-Stokes correlation. In Fig.~\ref{fig9}(a), the finite turn-on time of the acousto-optically switched write field smooths the rapid transient oscillations predicted for an abrupt turn-on in Figs.~\ref{fig2} and \ref{fig3}, yielding a temporal profile consistent with the measurement. After storage, the retrieved anti-Stokes wavepacket is divided into seven consecutive 200-ns temporal sections (I--VII). Because retrieval is driven by the classical read pulse, the wavepacket can be temporally gated at predetermined times without using Stokes-photon detection as a real-time feed-forward trigger.

Figure~\ref{fig9}(b) shows the integrated cross-correlations evaluated within the individual temporal sections, revealing the evolution of the correlation during readout of the stored spin wave. Here background counts from leakage light and detector dark counts account for approximately 30\% and 60\% of the detected counts in the Stokes and anti-Stokes channels, respectively, across the seven temporal sections. For the first temporal section, the theoretical cross-correlation decreases from approximately 250 in the absence of these backgrounds to approximately 100 when they are included, compared with the measured value of approximately 80. Despite this quantitative difference, the calculated correlation reproduces the overall trend across the seven temporal sections.

For comparison, Fig.~\ref{fig9}(c) shows the time-resolved SFWM cross-correlation evaluated with an 8-ns integration window. Further reducing the integration windows in either scheme produces little change in the extracted correlation values over the ranges considered. The DLCZ and SFWM configurations use the same experimental parameters, except that the driving and coupling pulses overlap temporally in SFWM. For SFWM, background counts from leakage light and detector dark counts account for approximately 10\% of the detected counts in both Stokes and anti-Stokes channels; in the absence of these backgrounds, the theoretical peak cross-correlation is approximately 75. The measured SFWM cross-correlation reaches a peak value of 57. The DLCZ cross-correlation exceeds this value through the first four temporal sections (800~ns), rising above 80 in the first two sections. This comparison shows that strong Stokes--anti-Stokes correlations can be retained under temporally gated DLCZ operation. Whereas temporal selection in continuous SFWM may require herald-conditioned modulation or post-selection, the intrinsic separation of the write and read stages in DLCZ-type SRS provides direct classical control of extended retrieval modes. These results demonstrate the potential of DLCZ-type SRS for temporally gated and time-bin-encoded photonic applications in high-dimensional quantum information processing.


\section*{DISCUSSION} \label{sec:Discussion}

We have presented a predictive open-system theory for DLCZ-type SRS and experimentally validated its key predictions. By combining Heisenberg--Langevin dynamics with Maxwell--Schr\"odinger propagation equations, the framework treats spontaneous emission, Langevin noise, spin-wave dynamics, and spatiotemporal propagation within a unified description. It predicts Stokes generation, retrieved anti-Stokes wavepackets, and time-resolved two-photon correlations. During the write process, the full time dependence of the mean atomic populations and optical coherence is retained, while the generated Stokes field and associated Raman coherences are treated to first order.

A central result is the distinction between DLCZ-type SRS and continuous SFWM in photon-generation and correlation dynamics. The write and read stages allow the phase-matched spin-wave excitation to accumulate, transiently enhancing Stokes generation. In continuous SFWM, this buildup is limited because the spin-wave coherence is coupled to anti-Stokes emission. During DLCZ retrieval, tuning coupling strength and detuning reshapes the anti-Stokes bandwidth, frequency, and temporal profile while largely preserving the normalized Stokes--anti-Stokes correlation.

Shortening the write pulse enhances the two-photon correlation because correlated coincidences scale approximately linearly with the mean spin-wave excitation number, whereas accidental backgrounds scale approximately quadratically. A separate classical read pulse combines write-stage correlation enhancement with controlled anti-Stokes retrieval. The temporal-slicing measurements further show that the retrieved wavepacket can be gated without Stokes-triggered feed-forward.

These properties support spectral and temporal matching across heterogeneous quantum-network nodes, including atomic quantum memories~\cite{QM1,QM2,QM3,QM4} and quantum frequency conversion modules~\cite{QFC1,QFC2,QFC3,QFC4,QFC5,QFC6,QFC7}. More broadly, atomic ensembles provide multimode capacity across temporal, spatial, and orbital-angular-momentum degrees of freedom~\cite{tbin1,tbin2,multiangle1,multiangle2,OAM1,OAM2,OAM3,OAM4}. The present framework therefore provides a quantitative basis for memory-compatible photon sources with controllable spectral, temporal, and correlation properties.


\section*{METHODS} \label{sec:method}

\FigTen

\subsection*{Numerical evaluation of the time-dependent propagator}

We evaluate the time-dependent write-stage dynamics by integrating the mean atomic equations on a two-segment temporal grid with step sizes of 1~ns for $T\leq1$~$\mu$s and 10~ns for $T>1$~$\mu$s. The temporal grid is further refined when required for a given parameter set to adequately resolve the early-time oscillations. The resulting populations and optical coherence determine $\alpha(T_k)$, $\beta(T_k)$, and $\mathbf{M}(s,T_k)$. The ordered propagator is constructed from $U_k=\exp[\mathbf{M}(s,T_k)\Delta T]$, with later-time matrices acting from the left according to Eq.~\eqref{eq11}. The inverse spatial Laplace transform is evaluated numerically via the Bromwich integral
\begin{equation}
	f(Z,T)=\frac{1}{2\pi}\int_{-\infty}^{\infty}du\,e^{(1/L+iu)Z}\,\widetilde f(1/L+iu,T),
	\label{eq24}
\end{equation}
using the trapezoidal rule along the contour $\mathrm{Re}(s)=1/L$. The spatial grid contains 100 points spanning $0.01L\leq Z\leq L$, while the Bromwich contour is sampled with 2001 points over $u\in[-2/Z_{\min},2/Z_{\min}]$, where $Z_{\min}=0.01L$. To avoid the removable singularity at $Z=0$ discussed in the analytic Green's functions, the numerical spatial grid begins at $Z=0.01L$ rather than $Z=0$; the contribution from the omitted interval was verified to be negligible in the spatial integrals. This procedure recovers the spatially dependent field and atomic operators over the numerical domain $0.01L\leq Z\leq L$.

Numerical convergence is verified by halving the temporal step sizes and, separately, doubling the number of Bromwich-contour quadrature points. Across the parameter range used in the figures, the Stokes generation rates and spin-wave profiles change by less than 0.1\% under either refinement test.


\subsection*{Undepleted large-detuning limit}

Although the time-dependent open-system treatment based on the piecewise-constant propagator retains the atomic, propagation, and noise dynamics, its expressions are not analytically transparent. For compact formulas with clearer physical interpretation, we consider the large-detuning regime, where the excited-state population remains small and the dynamics reduce to an effective Raman process. The expressions below also assume negligible accumulated population redistribution during the write interval. Large detuning and the undepleted-population approximation serve distinct roles. The former suppresses the instantaneous excited-state population and permits adiabatic elimination of the mean optical coherence, whereas the latter fixes the populations and renders the response coefficients time independent.

When population redistribution is neglected in this limit, the coefficients $\alpha$ and $\beta$ become time independent, and the coupling matrix $\mathbf{M}(s)$ depends only on the Laplace variable rather than on time. The propagators therefore depend only on the elapsed time interval, giving $\mathbf{W}(T_n,0)=e^{\mathbf{M}(s)T_n}$ and $\mathbf{w}(T_n,T') =e^{\mathbf{M}(s)(T_n-T')} =\mathbf{W}(T_n,T')$. These analytic Green's functions define the strict undepleted large-detuning limit and do not include the phenomenological replacement $\mathrm{OD}\rightarrow\mathrm{OD}_{\mathrm{eff}}(T)$ used for the blue curves in Fig.~\ref{fig2}. In this limit, Eq.~\eqref{eq15} describes the write-stage response before appreciable population transfer develops, and the corresponding spatiotemporal Green's functions take the analytic forms
\begin{align}
&\mathcal{G}_{11}^W(Z,T) \approx
e^{\frac{A-\gamma_{21}-\gamma_{23}}{4}T}I_0
\left(2\sqrt{\tfrac{\Gamma_{\rm R}TZ}{L}}\right),
\label{eq25}
\\
&\mathcal{G}_{12}^W(Z,T) \approx
\tfrac{i\Omega_d}{A}e^{-\frac{\gamma_{21}+\gamma_{23}}{4}T}
\nonumber\\
&\quad\times
\left[
e^{\frac{A}{4}T}I_0\left(2\sqrt{\tfrac{\Gamma_{\rm R}TZ}{L}}\right)
-e^{-\frac{A}{4}T}J_0\left(2\sqrt{\tfrac{\Gamma_{\rm R}TZ}{L}}\right)
\right],
\label{eq26}
\\
&\mathcal{G}_{21}^W(Z,T) \approx
\tfrac{i\Omega_d^*}{A}
\sqrt{\tfrac{\Gamma_{\rm R}T}{LZ}}
e^{-\frac{\gamma_{21}+\gamma_{23}}{4}T}
\nonumber\\
&\quad\times
\left[
e^{\frac{A}{4}T}I_1\left(2\sqrt{\tfrac{\Gamma_{\rm R}TZ}{L}}\right)
-e^{-\frac{A}{4}T}J_1\left(2\sqrt{\tfrac{\Gamma_{\rm R}TZ}{L}}\right)
\right],
\label{eq27}
\\
&\mathcal{G}_{22}^W(Z,T) \approx
-\sqrt{\tfrac{\Gamma_{\rm R}T}{LZ}}
e^{-\frac{A+\gamma_{21}+\gamma_{23}}{4}T}
J_1\left(2\sqrt{\tfrac{\Gamma_{\rm R}TZ}{L}}\right),
\label{eq28}
\end{align}
where $A=\sqrt{(\gamma_{21}-\gamma_{23})^2-4|\Omega_d|^2}$. The square-root branch is chosen continuously such that $A\rightarrow\gamma_{21}-\gamma_{23}$ as $|\Omega_d|\rightarrow0$. $J_{0,1}$ and $I_{0,1}$ denote ordinary and modified Bessel functions of the first kind, respectively. The effective Raman rate $\Gamma_{\rm R}=\Gamma\mathrm{OD}|\Omega_d|^2/(16\Delta_d^2)$ agrees with Eq.~\eqref{eq16}. The apparent $Z^{-1/2}$ factors in $\mathcal{G}_{21}^W$ and $\mathcal{G}_{22}^W$ remain finite as $Z\rightarrow0$ because $I_1(x),J_1(x)\propto x$ near the origin. For time-independent coefficients in this limit, the source-response functions satisfy $\mathcal{G}_{ij}^w(Z,T,T')\approx \mathcal{G}_{ij}^W(Z,T-T')$ for $i,j\in\{1,2\}$.

When population redistribution during the write interval is no longer negligible, the leading depletion effect can be incorporated phenomenologically into Eq.~\eqref{eq15} by replacing $\mathrm{OD}$, at each observation time $T$, with the time-dependent effective optical depth $\mathrm{OD}_{\mathrm{eff}}(T)=\mathrm{OD}\langle\hat\sigma_{11}(T)\rangle$ throughout the OD-dependent analytic response, including the explicit OD factors and the OD dependence of the Green's functions through $\Gamma_{\rm R}$. For the Langevin contribution $R_{s2}$, the diffusion matrix $\mathbf{D}(T')$ is additionally evaluated using the corresponding time-dependent quantities $\langle\hat\sigma_{11}(T')\rangle$, $\langle\hat\sigma_{33}(T')\rangle$, $\langle\hat\sigma_{31}(T')\rangle$, and $\langle\hat\sigma_{13}(T')\rangle$ in place of their strict undepleted values. The functional forms of the analytic Green's functions in Eqs.~\eqref{eq25}--\eqref{eq28} are retained under this quasistatic substitution; they are not replaced by the fully time-dependent Green's functions generated by the piecewise-constant propagator. These time-dependent mean values are obtained from $\mathbf{S}_0(T)=\exp(\mathbf{M}_0T)\mathbf{S}_0(0)$, where $\mathbf{S}_0=[\langle\hat{\sigma}_{11}\rangle,\langle\hat{\sigma}_{33}\rangle,\langle\hat{\sigma}_{31}\rangle,\langle\hat{\sigma}_{13}\rangle]^{\rm T}$ and
\begin{equation}
	\mathbf{M}_0
	=
	\begin{bmatrix}\begin{smallmatrix}
			0 & \Gamma/2 & -i\Omega_d/2 & i\Omega_d^*/2\\
			0 & -\Gamma & i\Omega_d/2 & -i\Omega_d^*/2\\
			-i\Omega_d^*/2 & i\Omega_d^*/2 & -(\Gamma+2i\Delta_d)/2 & 0\\
			i\Omega_d/2 & -i\Omega_d/2 & 0 & -(\Gamma-2i\Delta_d)/2
	\end{smallmatrix}\end{bmatrix}.
	\label{eq29}
\end{equation}
The initial conditions are $\langle\hat\sigma_{11}(0)\rangle=1$ and $\langle\hat\sigma_{33}(0)\rangle=\langle\hat\sigma_{31}(0)\rangle=\langle\hat\sigma_{13}(0)\rangle=0$.


\subsection*{Experimental details}

A cold $^{87}$Rb ensemble in a standard magneto-optical trap serves as the nonlinear medium. The effective level scheme comprises the two hyperfine ground states $|1\rangle=|5S_{1/2},F=1\rangle$ and $|2\rangle=|5S_{1/2},F=2\rangle$, together with the excited-state $|5P_{3/2},F=2\rangle$ manifold. The labels $|3\rangle$ and $|4\rangle$ denote the excited-state sublevels addressed by the write and read pathways, respectively, distinguishing the two Raman transitions in the effective model. Both driving and coupling fields address $\sigma^+$ transitions, thereby selecting the phase-matched $\sigma^+$-polarized Stokes and anti-Stokes modes collected experimentally. The three $F=1$ Zeeman sublevels form three double-$\Lambda$ pathways with the same topology but different Clebsch--Gordan (CG) coefficients~\cite{SteckRb87}.

In reducing this multilevel structure to the effective model, total spontaneous-decay rates are treated separately from couplings to the selected optical modes. Summing over all allowed emitted polarizations and final Zeeman sublevels gives equal branching fractions to the two ground-state manifolds, consistent with the rates specified in the main text. The CG dependence is incorporated into the effective Rabi frequencies and atom--field coupling constants. This separation retains the Zeeman-dependent transition strengths relevant to the selected write and read Raman pathways while accounting for emission into the remaining optical modes through the total spontaneous-decay rates.

The experimental energy-level diagrams and a schematic of the optical setup are shown in Fig.~\ref{fig10}(a) and Fig.~\ref{fig10}(b), respectively. At the atomic ensemble, the driving and coupling beams have $1/e^2$ intensity diameters of 250 and 310~$\mu$m, respectively. Details of the laser frequency stabilization are given in our previous work~\cite{JS2}. The generated Stokes and anti-Stokes photons are routed along separate optical paths, spectrally filtered by etalon filter sets with a bandwidth of approximately 100~MHz, and collected with an efficiency of approximately 2\% in each channel, before detection by single-photon counting modules. Their arrival times are recorded in list mode by a time-tagging multiscaler, allowing the same data set to be analyzed with different temporal windows for integrated cross-correlations and temporal slicing of the retrieved anti-Stokes wavepacket.

The experimental timing sequence is shown in Fig.~\ref{fig10}(c). The system operates at 400 Hz with a 2.5-ms cycle. Each cycle begins with 2.3 ms of laser cooling and optical pumping, followed by a 5-$\mu$s driving pulse with a continuous coupling field to induce SFWM~\cite{JS1,JS2,JS3}. The resulting correlated temporal profile provides an in situ OD calibration under the same alignment and atomic density as the subsequent DLCZ measurements. A 15-$\mu$s optical-pumping interval then prepares the population predominantly in $|1\rangle$ before the write pulse. The remaining cycle time is used for DLCZ-type SRS, so calibration and write--store--read measurements are performed within the same cycle. The AOM switching the driving field has approximately 35-ns turn-on and turn-off times, and the finite turn-on is included in the calculation for Fig.~\ref{fig9}(a).


\section*{DATA AVAILABILITY}

The data supporting the findings of this study are available from the corresponding author upon reasonable request.


\section*{CODE AVAILABILITY}

The custom numerical code used to generate the theoretical results of this study is available from the corresponding author upon reasonable request.



\section*{ACKNOWLEDGEMENTS}

This work was supported by the National Science and Technology Council of Taiwan under Grant Nos. 114-2112-M-006-007, 115-2119-M-007-004, and 115-2112-M-006-001. Additional support was provided by the Center for Quantum Science and Technology (CQST) through the Higher Education Sprout Project, which is funded by the Ministry of Education (MOE) in Taiwan.

\section*{AUTHOR CONTRIBUTIONS}

Y.-F.C. and J.-S.S. developed the theoretical framework. J.-S.S. and C.-W.L. performed the experiments. C.-M.Y. contributed to theoretical discussions, while I.A.Y. participated in discussions of the underlying physical mechanisms. J.-S.S. and Y.-F.C. analyzed the results and wrote the manuscript. All authors reviewed and approved the manuscript.

\section*{COMPETING INTERESTS}
The authors declare no competing interests.

\end{document}